\documentclass[11pt,a4paper,dvipdfmx]{article}

\usepackage{amsmath,amsthm,latexsym,amssymb}
\usepackage{graphicx,bm,fancybox,multirow,hhline,indentfirst}
\usepackage{ascmac,atbegshi}
\usepackage{natbib,verbatim,color}
\usepackage[colorlinks,citecolor=blue,urlcolor=blue,dvipdfmx]{hyperref}

\makeatletter
\def\section{\@startsection {section}{1}{\z@}{-3.5ex plus -1ex minus -.2ex}{2.3 ex plus .2ex}{\large\sc\centering}}
\def\subsection{\@startsection {subsection}{1}{\z@}{-3.5ex plus -1ex minus -.2ex}{2.3 ex plus .2ex}{\large}}
\makeatother

\theoremstyle{definition}

\newcommand{\argmin}{\mathop{\rm argmin}}
\def\E{{\rm E}}

\selectfont

\allowdisplaybreaks

\begin{document}

\begin{center}

\

{\Large\bf Smoothly varying ridge regularization}

\

{\large Daeju Kim$^{\rm a}$, Shuichi Kawano$^{\rm b}$, Yoshiyuki Ninomiya$^{\rm c}$\footnote{Corresponding author. Tel \& Fax: +81-50-5533-8527. {\it E-mail address}: ninomiya@ism.ac.jp (Y. Ninomiya).}}

\

{\small
{}$^{\rm a}${\it AI Strategy Office, Technology Unit, Softbank Corp.
\\
1-9-1 Higashi-Shimbashi, Minato-ku, Tokyo 105-7317, Japan}

{}$^{\rm b}${\it Graduate School of Informatics and Engineering, The University of Electro-Communications
\\
1-5-1 Chofugaoka, Chofu-shi, Tokyo 182-8585, Japan}

{}$^{\rm c}${\it Department of Statistical Inference and Mathematics, The Institute of Statistical Mathematics
\\
10-3 Midori-cho, Tachikawa-shi, Tokyo 190-8562, Japan}
}

\

\end{center}

\noindent {\bf Abstract}: 
A basis expansion with regularization methods is much appealing to the flexible or robust nonlinear regression models for data with complex structures.  When the underlying function has inhomogeneous smoothness, it is well known that conventional reguralization methods do not perform well.  In this case, an adaptive procedure such as a free-knot spline or a local likelihood method is often introduced as an effective method.  However, both methods need intensive computational loads.  In this study, we consider a new efficient basis expansion by proposing a smoothly varying regularization method which is constructed by some special penalties.  We call them adaptive-type penalties.  In our modeling, adaptive-type penalties play key rolls and it has been successful in giving good estimation for inhomogeneous smoothness functions.  A crucial issue in the modeling process is the choice of a suitable model among candidates.  To select the suitable model, we derive an approximated generalized information criterion (GIC).  The proposed method is investigated through Monte Carlo simulations and real data analysis.  Numerical results suggest that our method performs well in various situations.

\


\newpage

\section{Introduction}

Recently, nonlinear regression models with basis expansion have received considerable attention in various statistical and engineering fields. Basis expansion is widely used as an effective approach for data with complex structure. The essential idea behind basis expansion is to represent the underlying regression function as a linear combination of known nonlinear functions, which are called basis functions. In constructing the statistical model, various basis functions, such as natural cubic splines (\citealt{GreSil94}), $B$-splines (\citealt{Deb01}), and radial basis functions (\citealt{Pow81}) are used according to the structure of data or the purpose of analysis.
 Usually in basis expansion, we stop under-fitting by placing a lot of basis functions and prevent over-fitting by using regularization method for estimating their coefficients. 
 As the regularization, the ridge method (\citealt{HoeKen70}) or the lasso method (\citealt{Tib96}) is basically used.

While basis expansion with such a regularization method works well in many situations, it is often inappropriate when the underlying regression function has inhomogeneous smoothness. 
Let us call the region where the function is smoother the strongly smooth one and the region where the function is less smooth the weakly smooth one. 
Basis expansion that is described above often leads to under-fitting in the strongly smooth region and over-fitting in the weakly smooth region.
The local likelihood method (\citealt{FanGij96}, \citealt{Loa99}), defining a locally weighted log-likelihood by each explanatory variable and providing a predictor for the explanatory variable through minimizing it, can handle this problem.
Multivariate adaptive regression splines (MARS, \citealt{Fri91}), adaptive wavelet filtering (\citealt{DonJoh94}), adaptive sparse grids (\citealt{GarGT01}, \citealt{Heg02}) and free-knot splines (\citealt{DenMS98}, \citealt{DimGK01}, \citealt{MiyShe03}) can also handle it using a kind of optimal set of basis functions. %whose positions or shape have a kind of optimality.
 To find the optimal set, however, we must solve an optimiation problem with respect to the positions and shapes of the basis functions.
 In general, the objective function does not have tractable properties such as the unimodality.
 Therefore, in order to implement the optimization problem completely, it is necessary to compute the objective function for all the positions and shapes, which will be almost impossible if the number of basis functions is large.
 Needless to say, the local likelihood method also needs a huge amount of computations to be implemented completely because it requires an optimization by each point in the domain of function. 

The purpose of this study is to propose a simple new approach for the above-mentioned problem, without calculation by each explanatory variable or search for optimal basis functions.
To achieve our aim, we propose a new efficient nonlinear regression modeling with basis expansions.
The proposed method is based on the following idea.
To begin with, we prepare tuning parameters by each coefficient of the basis function, while conventional regularization methods usually have one or a few tuning parameters for the whole set of model's parameters.
The prepared many tuning parameters will lead over-fitting, and so we penalize them so that tuning parameters for basis functions whose positions are close have close values.
Then we obtain estimates of the coefficients and appropriate values of the tuning parameters simultaneously by maximizing the sum of the likelihood function, the penalty term for the coefficients and the penalty term for the tuning parameters.

We still have the problem of how penalize the tuning parameters.
To determine the weight of penalization, we put a coefficient for the penalty term, called hyper-tuning parameter.
Note that a larger value of the coefficient provides more weight of penalization.
Then the problem comes down to the selection of the hyper-tuning parameter, and its appropriate value can be obtained by minimizing an information criterion.
As the information criterion, we consider the generalized information criterion (GIC, \citealt{KonKit96}) because the estimator considered in this method is not the maximum likelihood one.
However, it is not possible to obtain an analytical expression of the GIC, because the estimation function for the proposed method is not expressed explicitly, unlike the estimation function of the ridge method.
We therefore propose an estimating function with explicit expression and derive an approximate GIC, which is assured by some asymptotics. 
Our proposed nonlinear regression modeling is investigated through some numerical examples and real data analysis.
Numerical results suggest that our proposed method performs well in various situations.

%Our procedure allows us to construct a statistical model which has some efficient properties: The degrees of smoothness are decided by using our proposed modeling so that they tend to be small on the strongly smooth regions and they tend to be large on weakly smooth regions.
%As a result, our procedure can avoid over-fitting and under-fitting from the estimated function on whole data region.
%Furthermore, one of the attractive features of our method is that it can be applied to surface estimation problems.
%For surface estimations, we use an information of first-order neighborhood of basis functions to estimate the several surfaces appropriately.

The rest of this article is organized as follows.
In Section \ref{sec2}, we describe a framework of Gaussian regression modeling based on basis expansion with the ridge method.
 Section \ref{sec3} studies a new nonlinear regression modeling by proposing a regularization method with adaptive-type penalties and derives the approximate GIC.
Section \ref{sec4} investigates the performance of our procedure by Monte Carlo simulations and real data analysis, while some concluding remarks are described in Section \ref{sec5}.

\section{Regression modeling via basis expansion}\label{sec2}

Let $\{ (\bm{x}_{i},y_{i}) \mid i=1,\ldots,n\}$ are $n$ sets of data obtained in terms of explanatory variables $\bm{x}\ (\in\mathbb{R}^p)$ and response variable $y\ (\in\mathbb{R})$. 
%\begin{align*}
%y_{i} = g(\bm{x}_{i}) + \varepsilon_{i}, \qquad i=1,\ldots,n,
%%\label{true} 
%\end{align*}
 We consider a nonlinear regression model based on basis expansion given by
\begin{align}
y_{i} = \bm{\beta}^{\rm T} \bm{\phi}(\bm{x}_{i}) + \varepsilon_{i} =  \sum_{j=1}^{m} \beta_{j} {\phi}_{j}(\bm{x}_{i}) + \varepsilon_{i}, \qquad i=1,\ldots,n.
\label{model}
\end{align}
Here, $\bm{\phi}(\bm{x}_{i}) = (\phi_{1}(\bm{x}_{i}), \phi_{2}(\bm{x}_{i}),\ldots,\phi_{m}(\bm{x}_{i}))^{\rm T}$ is a known basis function vector, $\bm{\beta} = (\beta_{1},\ldots,\beta_{m})^{\rm T}$ is its unknown coefficient vector, and $\varepsilon_{i}$ is an error term.
 Assuming that the error terms are independently and identically distributed according to ${\rm N}(0,\alpha)$ with unknown parameter $\alpha$, the log-likelihood function is $\sum_{i=1}^nl_i(\bm{\theta})$ from ($\ref{model}$), where
\begin{align}
l_i(\bm{\theta})=-\frac{1}{2}\log(2\pi\alpha)-\frac{1}{2\alpha}\{y_{i}-\bm{\beta}^{{\rm T}}\bm{\phi}(\bm{x}_{i})\}^2.
\label{likelihood}
\end{align}
 Here, $\bm{\theta} = (\alpha,\bm{\beta}^{\rm T})^{\rm T}$ is an unknown parameter vector to be estimated.

The maximum likelihood estimator often gives poor prediction accuracy, especially when the sample size $n$ is smaller than the number of basis functions $m$ (see, e.g., \citealt{KonKit08}).
Regularization method is one of the most popular and preferred methods to overcome this problem. It has been applied successfully to estimation problem in constructing flexible models with generalization framework.
 \cite{HoeKen70}, who proposed the $L_{2}$ norm of the penalty called the ridge, have successfully applied regularized regression in improving prediction performance through a bias variance trade-off. The ridge estimator can be obtained through minimizing the following regularized loss function:
\begin{align}
-2\sum_{i=1}^{n} l_i(\bm{\theta}) + n\lambda \bm{\beta}^{\rm T} \bm{\beta},
\label{loss}
\end{align}
where $\lambda$ is a non-negative tuning parameter that balances between goodness of fit and model complexity.
 The ridge estimator $\hat{\bm{\theta}}=(\hat{\alpha},\hat{\bm{\beta}}^{\rm T})^{\rm T}$ of $\bm{\theta}=(\alpha,\bm{\beta}^{\rm T})^{\rm T}$ is expressed by
\begin{align*}
\hat{\alpha} = \frac{1}{n} \sum_{i=1}^n \{y_i - \hat{\bm{\beta}}^{\rm T} \bm{\phi}(\bm{x}_i)\}^2 \quad {\rm and} \quad \hat{\bm{\beta}}=\bigg\{\sum_{i=1}^n \bm{\phi}(\bm{x}_i)\bm{\phi}(\bm{x}_i)^{\rm T} +n \lambda \hat{\alpha} \bm{I}_{m}\bigg\}^{-1} \sum_{i=1}^n \bm{\phi}(\bm{x}_i) y_i, 
%\label{ridge}
\end{align*}
where $\bm{I}_m$ is the $m$-dimensional identity matrix.
 Note that $\hat{\alpha}$ and $\hat{\bm{\beta}}$ depend on each other. Therefore, we provide an appropriate initial value for $\hat{\bm{\beta}}$ first, then $\hat{\alpha}$ and $\hat{\bm{\beta}}$ are updated until convergence.

In order to complete the formulation of the ridge method, we need to determine an appropriate value of the tuning parameter $\lambda$. 
This determination can be viewed as a model selection problem, and here we regard the minimizer of so-called GIC proposed by \citet{KonKit96} as the optimal value of $\lambda$.
The GIC is an AIC-type information criterion generalized for $M$-estimation methods, and actually this is an asymptotically unbiased estimator of twice the Kullback-Leibler divergence (\citealt{KulLei51}) between the true distribution and the estimated distribution minus some constant. 
Let $\sum_{i=1}^{n}\bm{\psi}_i({\bm{\theta}})$ and $\sum_{i=1}^{n}l_i({\bm{\theta}})$ be respectively an estimating function and a log-likelihood function with respect to a parameter vector $\bm{\theta}$, where $\bm{\psi}_i(\cdot)$ and $l_i(\cdot)$ are functions depending on the $i$-th data. 
Then the GIC is provided by
\begin{align}
-2\sum_{i=1}^{n}l_i(\hat{\bm{\theta}})+2{\rm tr}\bigg\{ 
{\rm E}\bigg(-\sum_{i=1}^{n}\frac{\partial\bm{\psi}_i}{\partial\bm{\theta}^{\rm T}}\bigg|_{\bm{\theta}=\bm{\theta}^*}\bigg)^{-1}
{\rm E}\bigg(\sum_{i=1}^{n}\bm{\psi}_i\frac{\partial l_i}{\partial\bm{\theta}^{\rm T}}\bigg|_{\bm{\theta}=\bm{\theta}^*}\bigg)
\bigg\} \bigg|_{\bm{\theta}^*=\hat{\bm{\theta}}},
\label{GeneralGIC}
\end{align}
where $\hat{\bm{\theta}}$ and $\bm{\theta}^*$ are respectively the $M$-estimator satisfying $\sum_{i=1}^{n}\bm{\psi}_i(\hat{\bm{\theta}})=\bm{0}$ and its limit satisfying ${\rm E}\{\sum_{i=1}^{n}\bm{\psi}_i(\bm{\theta}^*)\}=\bm{0}$. 
 When these expectations cannot be calculated completely, we have only to replace the left expectations by their empirical versions and use the replaced one as the GIC.  
 For example,  
\begin{align*}
-2\sum_{i=1}^{n}l_i(\hat{\bm{\theta}})+2{\rm tr}\bigg\{
\bigg(-\sum_{i=1}^{n}\frac{\partial\bm{\psi}_i}{\partial\bm{\theta}^{\rm T}}\bigg|_{\bm{\theta}=\bm{\theta}^*}\bigg)^{-1}
\bigg(\sum_{i=1}^{n}\bm{\psi}_i\frac{\partial l_i}{\partial\bm{\theta}^{\rm T}}\bigg|_{\bm{\theta}=\bm{\theta}^*}\bigg)
\bigg\}
%\label{GeneralGIC2}
\end{align*}
is a simple candidate to use, but it is better to calculate the expectations whenever possible in order to stabilize the criterion. 

For the above-mentioned ridge method in the Gaussian linear regression with basis expansion, $l_i({\bm{\theta}})$ is defined in (\ref{likelihood}), and $\sum_{i=1}^n\bm{\psi}_i({\bm{\theta}})$ is the derivative of (\ref{loss}), that is, $\bm{\psi}_i({\bm{\theta}})=-2\partial l_i(\bm{\theta})/\partial\bm{\theta}+2\lambda(0,\bm{\beta}^{\rm T})^{\rm T}$. 
 In this setting, the expectations in (\ref{GeneralGIC}) can be calculated to some extent. 
 Actually, letting $\varepsilon_i^*\equiv y_i-\bm{\beta}^{*{\rm T}}\bm{\phi}(\bm{x}_i)$, it holds
\begin{align*}
\E\bigg(-\sum_{i=1}^n\frac{\partial\bm{\psi}_i}{\partial\bm{\theta}^{\rm T}}\bigg|_{\bm{\theta}=\bm{\theta}^*}\bigg)=
\begin{pmatrix}
\displaystyle \frac{n}{\alpha^{*2}}
%\frac{1}{\alpha^{*3}}\{y_i-\bm{\beta}^{*{\rm T}}\bm{\phi}(\bm{x}_i)\}^2-\frac{1}{2\alpha^{*2}} 
& 
\displaystyle \frac{2}{\alpha^{*2}}\sum_{i=1}^n\E(\varepsilon_i^*)\bm{\phi}(\bm{x}_i)^{\rm T}
\\[1.5ex]
\displaystyle \frac{2}{\alpha^{*2}}\sum_{i=1}^n\E(\varepsilon_i^*)\bm{\phi}(\bm{x}_i)
&
\displaystyle \frac{2}{\alpha^{*}}\sum_{i=1}^n\bm{\phi}(\bm{x}_i)\bm{\phi}(\bm{x}_i)^{\rm T}+2n\lambda\bm{I}_m
\end{pmatrix}
\end{align*}
and
\begin{align*}
{\rm E}\bigg(\sum_{i=1}^n\bm{\psi}_i\frac{\partial l_i}{\partial\bm{\theta}^{\rm T}}\bigg|_{\bm{\theta}=\bm{\theta}^*}\bigg)
=\begin{pmatrix}
\displaystyle \frac{1}{2\alpha^{*4}}\sum_{i=1}^n\E(\varepsilon_i^{*4})-\frac{n}{2\alpha^{*2}}
% & \displaystyle \frac{1}{\alpha^3}\E[\varepsilon_i\{y_i-\bm{\beta}^{\rm T}\bm{\phi}(\bm{x}_i)\}^2]\bm{\phi}(\bm{x}_i)^{\rm T}
& \displaystyle \sum_{i=1}^n\E\bigg(\frac{1}{\alpha^{*3}}\varepsilon_i^{*3}-\frac{1}{\alpha^{*2}}\varepsilon_i^*\bigg)\bm{\phi}(\bm{x}_i)^{\rm T}
\\[1.5ex]
% \displaystyle \frac{1}{\alpha^3}\E[\varepsilon_i\{y_i-\bm{\beta}^{\rm T}\bm{\phi}(\bm{x}_i)\}^2]\bm{\phi}(\bm{x}_i)
\displaystyle \sum_{i=1}^n\E\bigg(\frac{1}{\alpha^{*3}}\varepsilon_i^{*3}-\frac{1}{\alpha^{*2}}\varepsilon_i^*\bigg)\bm{\phi}(\bm{x}_i)
& \displaystyle \frac{2}{\alpha^{*2}}\sum_{i=1}^n\E(\varepsilon_i^2)\bm{\phi}(\bm{x}_i)\bm{\phi}(\bm{x}_i)^{\rm T}
% & \displaystyle \frac{1}{\alpha}(y_i-\bm{\beta}^{\rm T}\bm{\phi}(\bm{x}_i))\bm{\phi}(\bm{x}_i) \bigg\{\frac{2}{\alpha}(y_i-\bm{\beta}^{\rm T}\bm{\phi}(\bm{x}_i))\bm{\phi}(\bm{x}_i)^{\rm T}-2\tilde{\Lambda}\bm{\beta}-\frac{\partial\bm{\tilde{\lambda}}^{\rm T}}{\partial\bm{\beta}}({\cal B}^2\bm{1}+2\gamma_1D\tilde{\bm{\lambda}}+\gamma_2\tilde{\Lambda}^{-1}\bm{1})\bigg\}
\end{pmatrix},
\end{align*}
and therefore we obtain the GIC based on these expectations.
 When $\alpha$ is known, we can see from these expectations that half the penalty term in the GIC reduces to 
\begin{align*}
{\rm tr}\bigg[\bigg\{\sum_{i=1}^n\bm{\phi}(\bm{x}_i)\bm{\phi}(\bm{x}_i)^{\rm T}+n\lambda\alpha\bm{I}_m\bigg\}^{-1}\bigg\{\sum_{i=1}^n\bm{\phi}(\bm{x}_i)\bm{\phi}(\bm{x}_i)^{\rm T}\bigg\}\bigg],
\end{align*}
which is known as the effective degree of the freedom for the ridge method (see, e.g., Chapter 5 in \citealt{HasTF09}). 

\begin{figure}[t]
\begin{center}
\vspace{-5mm}
\includegraphics[width=8cm]{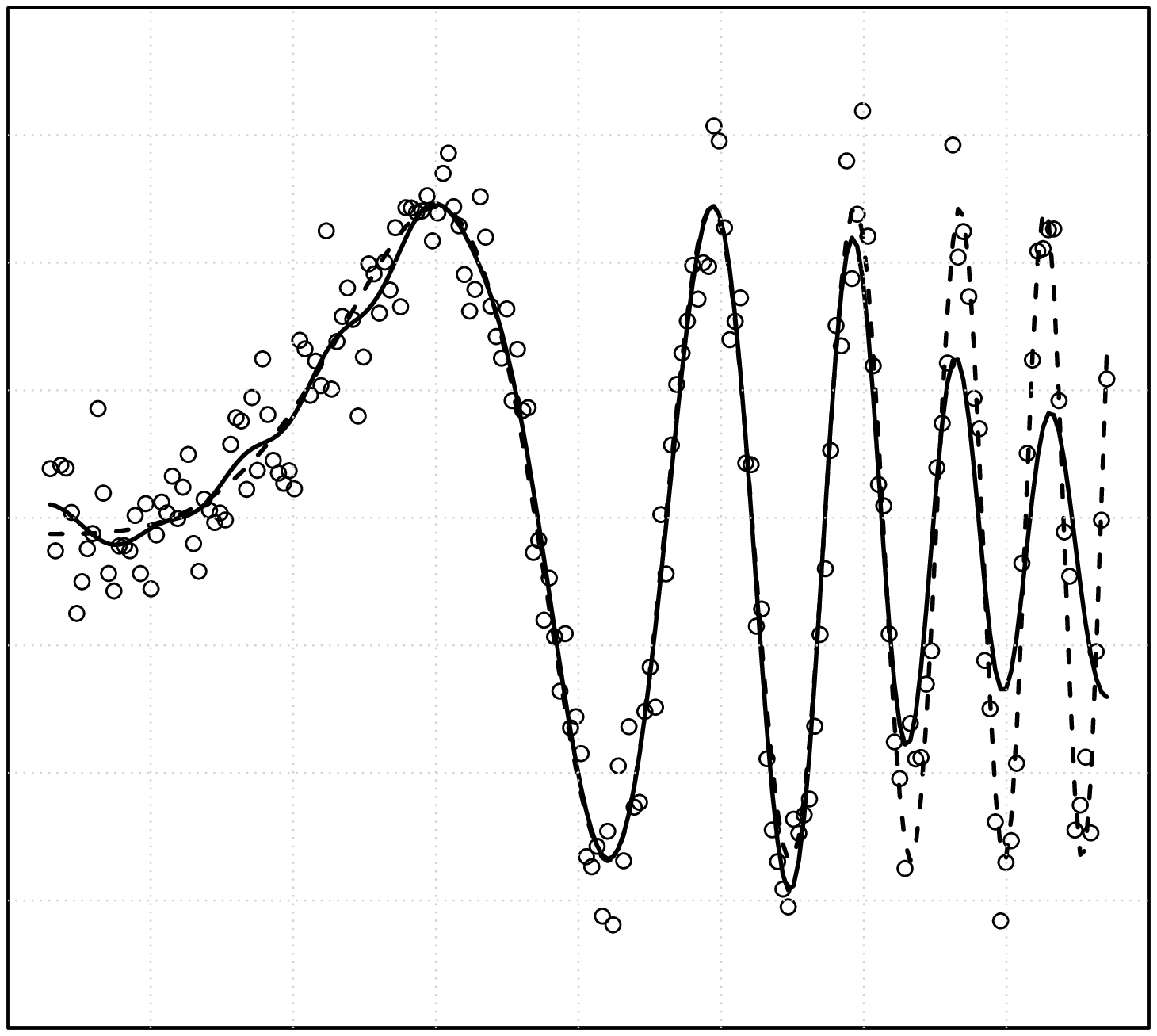}
\includegraphics[width=8cm]{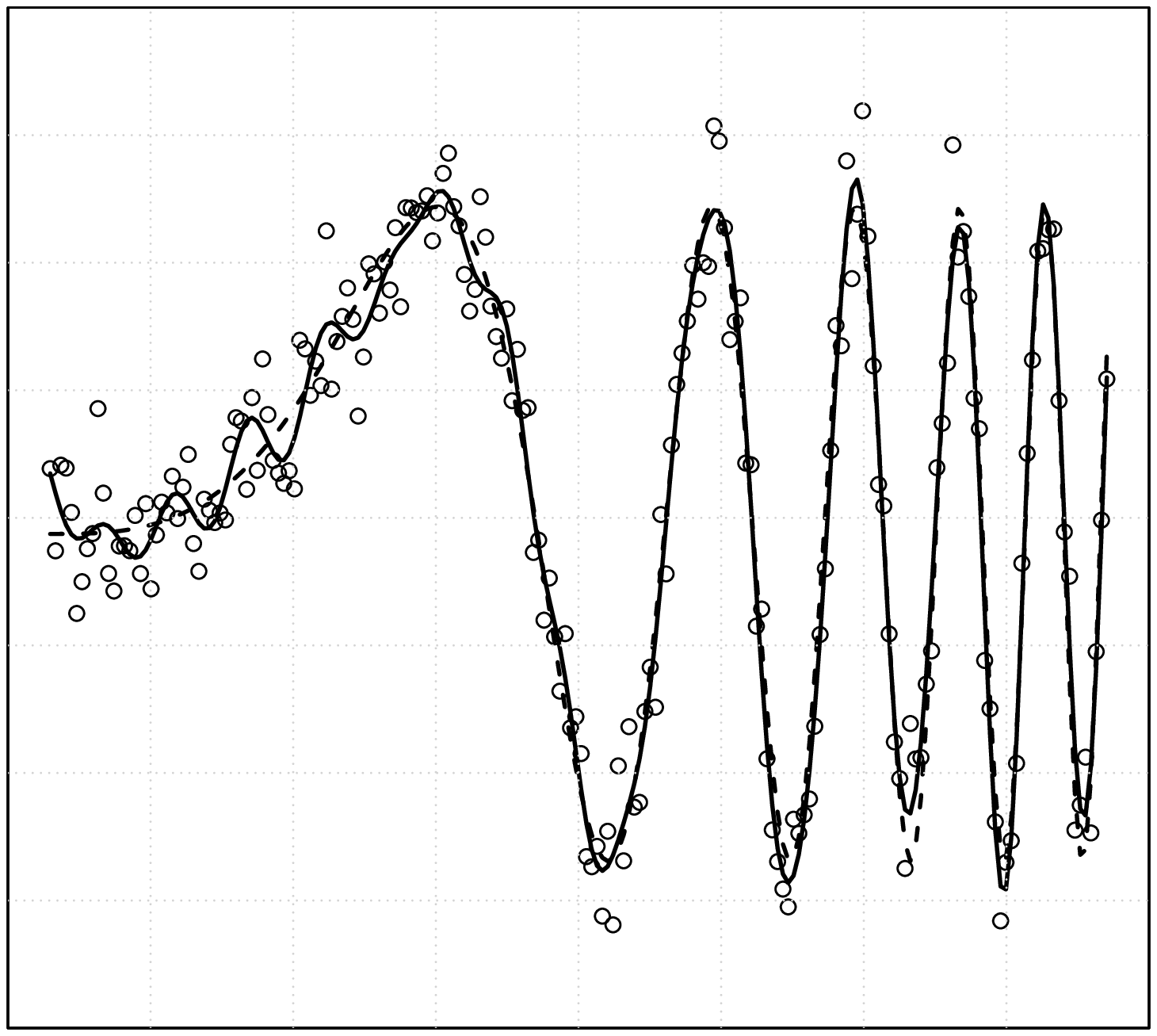}
\vspace{-10mm}
\caption{Estimated curves based on conventional basis expansion with a strong regularization (left) and a weak regularization (right). Solid lines, broken lines, and dots depict the estimated curves, the true regression functions, and the data, respectively.}
\label{Fig1}
\end{center}
\end{figure}

When the underlying regression function has inhomogeneous smoothness, conventional regularization methods often lead over-fitting at strongly smooth regions or/and under-fitting at weakly smooth regions.
 See estimated curves by the ridge method in Figure \ref{Fig1} as an illustrative example. 
 We observe the under-fitting in the left panel imposing on a strong regularization and the over-fitting in the right panel imposing on a weak regularization.   
 The local likelihood method is a tool to overcome this problem.
 For each point $\bm{x}_i\in\mathbb{R}^p$ ($i=1,\ldots,n$), it assigns large weights to observations close to $\bm{x}_i$, constructs a local likelihood based on the weights and obtains estimators by maximizing it. 
 Multivariate adaptive regression splines, adaptive wavelet filtering, adaptive sparse grids and free-knot splines can also overcome the problem by selecting the positions or/and shapes of the basis functions $\bm{\phi}(\bm{x})=(\phi_1(\bm{x}),\ldots,\phi_m(\bm{x}))^{\rm T}$ data-adaptively.
 However, such a local likelihood and such adaptive methods need intensive computational loads, which will be almost impossible to be implemented completely if the data size $n$ or the number of basis functions $m$ is large.

\section{Proposed method}\label{sec3}

\subsection{Statistical Setup}\label{subsec31}

Let $\{(\bm{x}_i,y_i)\ |\ i=1,\ldots,n\}$ be $n$ sets of data obtained in terms of explanatory variables $\bm{x}$ and response variable $y$.
We assume that the behavior of $y$ is characterized by an unknown parameter vector $\bm{\alpha}=(\alpha_1,\ldots,\alpha_{\ell})$ and a linear combination $\bm{\beta}^{\rm T}\bm{\phi}(\bm{x})$ with an unknown coefficient vector $\bm{\beta}=(\beta_1,\ldots,\beta_m)$ and a vector of known functions $\bm{\phi}(\cdot)=(\phi_1(\cdot),\ldots,\phi_m(\cdot))$.
For measuring the goodness of fit, we use a loss function $L\{y,\bm{\alpha},\bm{\beta}^{\rm T}\bm{\phi}(\bm{x})\}$.
For this setting, let us consider the case where the estimation of $\bm{\alpha}$ and $\bm{\beta}$ obtained by minimizing $\sum_{i=1}^nL\{y_i,\bm{\alpha},\bm{\beta}^{\rm T}\bm{\phi}(\bm{x}_i)\}$ leads to over-fitting. 
For example, it often occurs when $n$ is not large compared to $m$. 
Regularization methods are popular and preferred to overcome this problem, and first we consider a regularization method which provides the following type of estimators:
\begin{align*}
\argmin_{(\bm{\alpha},\bm{\beta})}\bigg[\sum_{i=1}^nL\{y_i,\bm{\alpha},\bm{\beta}^{\rm T}\bm{\phi}(\bm{x}_i)\}
+n\lambda\sum_{j=1}^mr(\beta_j)\bigg],
\end{align*}
where $r(\cdot)$ is a positive-value penalty function and $\lambda$ is a tuning parameter.
Note that the regularization by putting $\lambda\bm{\beta}^{\rm T}K\bm{\beta}$ for a positive definite matrix $K$ can be included in this formulation by reparametrization. 

This estimation method is required in a wide class of modern statistical problems.
Letting $g\{\bm{\alpha},\bm{\beta}^{\rm T}\bm{\phi}(\bm{x})\}$ be a predictor for $y$, for example, $L(y,s,t)=(y-t)^2$ and $g(u,v)=v$ are used in regression analysis with linear basis expansion models, $L(y,s,t)=-yt+\log(1+e^t)$ and $g(u,v)=\{{\rm sign}(v)+1\}/2$ are used in discriminant analysis with nonparametric logistic regression model. 
In the context of support vector machine, $L(y,s,t)=(|y-t|-\epsilon)_+$ and $g(u,v)=v$ for a positive value $\epsilon$ or $L(y,s,t)=(1-yt)_+$ and $g(u,v)={\rm sign}(v)$ are used in regression or discriminant analysis, respectively. 
See, e.g., Capters 5 and 12 in \cite{HasTF09} and Chapter 7 in \cite{Bis06}.

\subsection{Estimation}\label{subsec32}

In this section, we first describe the basic idea behind our proposed method in a general framework. 
 The main aim of this approach is to appropriately estimate the underlying regression function which has inhomogeneous smoothness and less the computational load. 
 To achieve our aim, we propose a new nonlinear regression modeling with smoothly varying regularization method as follows:
\begin{align}
\sum_{i=1}^{n}\bigg[L\{y_{i},\bm{\alpha},\bm{\beta}^{\rm T}\bm{\phi}(\bm{x}_{i})\}+\sum_{j=1}^{m}\lambda_{j}r(\beta_{j})+\gamma_{1}\sum_{j=2}^{m}(\lambda_{j}-\lambda_{j-1})^2-\gamma_{2}\sum_{j=1}^{m}\log\lambda_{j}\bigg],
\label{propose}
\end{align}
where $\bm{\lambda} = (\lambda_{1},\ldots,\lambda_{m})^{\rm T}$ is an $m$-dimensional tuning parameter vector, 
%$\Lambda = {\rm diag}(\bm{\lambda})$, 
and $\gamma_{1}$ and $\gamma_{2}$ are hyper-tuning parameters.

In our approach, the proposed adaptive-type penalties in ($\ref{propose}$) play key roles: The first penalty is to assign different tuning parameters to different coefficients, which promotes adaptability of tuning parameters. The second penalty is a sum of squared values of differences for tuning parameters, and encourages their continuity. The third penalty is a sum of logarithmic values of tuning parameters which keep the values of tuning parameters from being shrunk toward zero. 
 By evaluating $\bm{\lambda}$ which gives a small value of the regularized log-likelihood function in ($\ref{propose}$), we can obtain smoothly varying tuning parameters.
 They can be expected to impose on a strong and weak regularizations at strongly and weakly smooth regions, respectively.  
 The hyper-tuning parameters $\gamma_{1}$ and $\gamma_{2}$ in ($\ref{propose}$) control the magnitude of effects for the second and third penalties. At the smaller value of $\gamma_{1}$, the tuning parameters tend to be decided freely, and then they tend to be continuous at larger value of $\gamma_{1}$. At the smaller value of $\gamma_{2}$, they tend to take a value which close to zero, although they tend to be large at the larger value of $\gamma_{2}$.

If the tuning parameters $\gamma_{1}$ and $\gamma_{2}$ are given, estimators of $\bm{\alpha}$ and $\bm{\beta}$ and an appropriate value of $\bm{\lambda}$ can be obtained by minimizing the proposed function in ($\ref{propose}$) with respect to $\bm{\alpha}$, $\bm{\beta}$ and $\bm{\lambda}$. 
We denote them by $\hat{\bm{\alpha}}$, $\hat{\bm{\beta}}$ and $\hat{\bm{\lambda}}$, which satisfy
\begin{align*}
(\hat{\bm{\alpha}},\hat{\bm{\beta}})=\argmin_{(\bm{\alpha},\bm{\beta})}\bigg(\sum_{i=1}^n\bigg[L\{y_i,\bm{\alpha},\bm{\beta}^{\rm T}\bm{\phi}(\bm{x}_{i})\}+\sum_{j=1}^{m}\hat{\lambda}_{j}r(\beta_{j})\bigg]\bigg)
\end{align*}
and
\begin{align*}
\hat{\lambda}_{j}=(2\gamma_1\hat{\lambda}_{j+1}+2\gamma_1\hat{\lambda}_{j-1}-r(\hat{\beta}_{j})+[\{2\gamma_1\hat{\lambda}_{j+1}+2\gamma_1\hat{\lambda}_{j-1}-r(\hat{\beta}_{j})\}^2+16\gamma_1\gamma_2]^{1/2})/(8\gamma_{1}).
\end{align*}
Since the solutions depend on each other, an iterative procedures are required. 
To get $\hat{\bm{\alpha}}$, $\hat{\bm{\beta}}$ and $\hat{\bm{\lambda}}$, we use the following numerical estimation algorithm.
\begin{description}
\item{{\bf Step 1 }} Set the values of hyper-tuning parameters $\gamma_{1}$ and $\gamma_{2}$.
\item{{\bf Step 2 }} Give an appropriate initial value of the tuning parameters, $\hat{\bm{\lambda}}^{(0)}=(\hat{\lambda}_1^{(0)},\ldots,\hat{\lambda}_m^{(0)})$, such as the optimal value of the tuning parameter for the naive ridge method.
\item{{\bf Step 3 }} Substitute $0$ for $t$.
\item{{\bf Step 4 }} Give $(\hat{\bm{\alpha}}^{(t+1)},\hat{\bm{\beta}}^{(t+1)})$ by $\argmin_{\bm{\alpha},\bm{\beta}}(\sum_{i=1}^n[L\{y_i,\bm{\alpha},\bm{\beta}^{\rm T}\bm{\phi}(\bm{x}_{i})\}+\sum_{j=1}^{m}\hat{\lambda}_{j}^{(t)}r(\beta_{j})])$.
\item{{\bf Step 5 }}
Set $\hat{\lambda}_{0}^{(t+1)}=\hat{\lambda}_{m+1}^{(t+1)}=0$ and give $\hat{\lambda}_{j}^{(t+1)}$ by $(2\gamma_1\hat{\lambda}_{j+1}^{(t)}+2\gamma_1\hat{\lambda}_{j-1}^{(t+1)}-r(\hat{\beta}_{j}^{(t+1)})+[\{2\gamma_1\hat{\lambda}_{j+1}^{(t)}+2\gamma_1\hat{\lambda}_{j-1}^{(t+1)}-r(\hat{\beta}_{j}^{(t+1)})\}^2+16\gamma_1\gamma_2]^{1/2})/(8\gamma_{1})$ for $j=1,\ldots,m$. 
\item{{\bf Step 6 }} Substitute $t+1$ for $t$.
\item{{\bf Step 7 }} Repeat Steps 4, 5 and 6 until the values converge.
\end{description}
Note that $L(\cdot)$ and $r(\cdot)$ are convex in many cases and so Step 4 can be easily conducted by using Newton-Raphson method.

\subsection{Model selection}\label{subsec33}

Our proposed regularization method depends on hyper-tuning parameters $\gamma_{1}$ and $\gamma_{2}$, and it is indispensable to determine their appropriate values. 
Then we consider the GIC introduced in Section \ref{sec2} for our method. 
In our framework, however, it is difficult to directly derive an analytical form of the GIC. 
Therefore, here we focus on the Gaussian linear regression setting with a ridge-type regularization and approximate the GIC based on some asymptotics. 
Concretely, we suppose that the loss function is twice the negative log-likelihood, that is, 
\begin{align*}
L\{y_i,\alpha,\bm{\beta}^{\rm T}\bm{\phi}(\bm{x}_i)\}=-2l_i(\bm{\theta})=\log(2\pi\alpha)+\{y_i-\bm{\beta}^{\rm T}\bm{\phi}(\bm{x}_i)\}^2/\alpha, 
\end{align*}
and the penalty function is $r(\beta_j)=\beta_j^2$. 
For the other settings, we recommend to naively use the cross-validation (\citealt{Sto74}) by overlooking its computational costs.

In this setting, the estimation function is $\sum_{i=1}^n\bm{\psi}_i(\bm{\theta})$, where $\bm{\theta}=(\alpha,\bm{\beta}^{\rm T})^{\rm T}$ and  
\begin{align}
\bm{\psi}_i(\bm{\theta})=-\frac{\partial}{\partial\bm{\theta}}\min_{\bm{\lambda}}\bigg[
%-\frac{n}{2}\log(2\pi\alpha)-\frac{1}{2\alpha}\sum_{i=1}^{n}(y_i-\bm{\beta}^{\rm T}\bm{\phi}(\bm{x}_i))^2
L\{y_i,\alpha,\bm{\beta}^{\rm T}\bm{\phi}(\bm{x}_i)\}
+\sum_{j=1}^m\lambda_j\beta_j^2+\gamma_1\sum_{j=2}^m(\lambda_j-\lambda_{j-1})^2-\gamma_2\sum_{j=1}^m\log\lambda_j\bigg].
\label{proposed_est_func}
\end{align}
We can obtain an AIC-type information criterion from this estimation function, but it is difficult to express the criterion exactly because the estimation function depends on the estimated tuning parameter vector $\hat{\bm{\lambda}}$, which is the minimizer of $\bm{\lambda}$ appearing in ($\ref{proposed_est_func}$). 
% \begin{align}
% \arg \max_{\bm{\beta}, \sigma^2} \left\{\arg \max_{\bm{\lambda} \in \mathbb{R}^{m}} \left\{ \sum_{i=1}^{n} \log f(y_{i}|\bm{x}_{i};\bm{\beta},\sigma^2) - \frac{1}{2} \bm{\beta}^{\rm T} \Lambda \bm{\beta} -\frac{\gamma_{1}}{2}  \bm{\lambda}^{\rm T} D \bm{\lambda} + \frac{\gamma_{2}}{2} \log \det (\Lambda) \right\} \right\}.
% \end{align}
 Here, we propose an estimation function using an approximate vector $\tilde{\bm{\lambda}}=(\tilde{\lambda}_1,\ldots,\tilde{\lambda}_p)$ instead of $\hat{\bm{\lambda}}$ in ($\ref{proposed_est_func}$).
 We utilize a simple approximation as follows:
 First, to get the optimal value of $\bm{\lambda}$, we consider the derivative of ($\ref{propose}$) with respect to $\lambda_{j}$,
\begin{align}
\beta_{j}^2 + 2 \gamma_{1}(\lambda_{j+1}-2 \lambda_{j} + \lambda_{j-1}) - \gamma_{2}/\lambda_{j}. 
\label{dev_pro}
\end{align}
Checking for zero of ($\ref{dev_pro}$), we have the equation,
\begin{align}
{\lambda}_{j} = \frac{\gamma_{2}}{\beta_{j}^2 + 2 \gamma_{1}(\lambda_{j+1} -2\lambda_{j} +\lambda_{j-1} )}. \label{dev_pro2}
\end{align}
When we consider $\gamma_{1}=0$, it leads to ${\lambda}_{j} = \gamma_{2}/\beta_{j}^2$.
 By substituting it into ${\lambda}_{j}$ in the right hand of ($\ref{dev_pro2}$), we get the following approximation,
\begin{align*}
\tilde{\lambda}_{j} = \frac{\gamma_{2}}{ {\beta_{j}^{2} + 2\gamma_{1}(\gamma_{2}/\beta_{j+1}^2 - 2\gamma_{2}/\beta_{j}^2 + \gamma_{2}/\beta_{j-1}^2)}}.
\end{align*}
As a result, the estimation function in ($\ref{proposed_est_func}$) is approximated by
\begin{align}
\tilde{\bm{\psi}}_i({\bm{\theta}}) = -\frac{\partial}{\partial \bm{\theta}} [ L\{y_i,\alpha,\bm{\beta}^{\rm T}\bm{\phi}(\bm{x}_i)\} + \bm{\beta}^{\rm T} \tilde{\bm{\Lambda}} \bm{\beta} + \gamma_{1} \tilde{\bm{\lambda}}^{\rm T} \bm{D} \tilde{\bm{\lambda}} - \gamma_{2} \log \det (\tilde{\bm{\Lambda}}) ], 
\label{proposed_est_func_app}
\end{align}
where 
\begin{align*}
\tilde{\bm{\Lambda}}={\rm diag}(\tilde{\bm{\lambda}})\equiv
\begin{pmatrix}
\tilde{\lambda}_1&0&\cdots&\cdots&0\\
0&\tilde{\lambda}_2&\ddots&&\vdots\\
\vdots&\ddots&\ddots&\ddots&\vdots\\
\vdots&&\ddots&\tilde{\lambda}_{p-1}&0\\
0&\cdots&\cdots&0&\tilde{\lambda}_p
\end{pmatrix}
\quad{\rm and}\quad 
\bm{D}\equiv\begin{pmatrix}
1&-1&0&\cdots&\cdots&\cdots&0
\\-1&2&-1&\ddots&&&\vdots
\\0&-1&2&\ddots&\ddots&&\vdots
\\\vdots&\ddots&\ddots&\ddots&\ddots&\ddots&\vdots
\\\vdots&&\ddots&\ddots&2&-1&0
\\\vdots&&&\ddots&-1&2&-1
\\0&\cdots&\cdots&\cdots&0&-1&1
\end{pmatrix}.
\end{align*}   
Letting $\bm{S}\equiv2(\partial\tilde{\bm{\lambda}}^{\rm T}/\partial\bm{\beta})^{-1}{\rm diag}(\bm{\beta})+2{\rm diag}(\bm{\beta})(\partial\tilde{\bm{\lambda}}/\partial\bm{\beta}^{\rm T})^{-1}+2\gamma_1\bm{D}+\gamma_2\tilde{\Lambda}^{-2}$, $t_j\equiv\beta_j^2+2\gamma_1(\bm{D}\tilde{\bm{\lambda}})_j-\gamma_2/\tilde{\lambda}_j$, $\bm{u}\equiv{\rm diag}(\bm{\beta})^2\bm{1}_m+2\gamma_1\bm{D}\tilde{\bm{\lambda}}+\gamma_2\tilde{\bm{\Lambda}}^{-1}\bm{1}_m$ and $\varepsilon_i^*\equiv y_i-\bm{\beta}^{*\rm T}\bm{\phi}(\bm{x}_i)$, where $\bm{1}_m$ is the $m$-dimensional $1$-vector, we have 
\begin{align*}
-\frac{\partial\tilde{\bm{\psi}}_i}{\partial\bm{\theta}^{\rm T}}\bigg|_{\bm{\theta}=\bm{\theta}^*}
=\begin{pmatrix}
\displaystyle \frac{2}{\alpha^{*3}}\varepsilon_i^{*2}-\frac{1}{\alpha^{*2}}
& \displaystyle \frac{2}{\alpha^{*2}}\varepsilon_i^*\bm{\phi}(\bm{x}_i)^{\rm T}
\\[1.5ex]
\displaystyle \frac{2}{\alpha^{*2}}\varepsilon_i^*\bm{\phi}(\bm{x}_i)
& \displaystyle \frac{2}{\alpha^*}\bm{\phi}(\bm{x}_i)\bm{\phi}(\bm{x}_i)^{\rm T}+\bigg(2\tilde{\bm{\Lambda}}+\frac{\partial\tilde{\bm{\lambda}}^{\rm T}}{\partial\bm{\beta}}
%\bigg\{2\bigg(\frac{\partial\tilde{\bm{\lambda}}^{\rm T}}{\partial\bm{\beta}}\bigg)^{-1}{\rm diag}(\bm{\beta})+2{\rm diag}(\bm{\beta})\bigg(\frac{\partial\tilde{\bm{\lambda}}}{\partial\bm{\beta}^{\rm T}}\bigg)^{-1}+2\gamma_1D+\gamma_2\tilde{\Lambda}^{-2}\bigg\}
\bm{S}\frac{\partial\tilde{\bm{\lambda}}}{\partial\bm{\beta}^{\rm T}}
+\sum_{j=1}^p
%\bigg\{\beta_j^2+2\gamma_1(D\tilde{\bm{\lambda}})_j-\frac{\gamma_2}{\tilde{\lambda}_j}\bigg\}
t_j\frac{\partial^2\tilde{\lambda}_j}{\partial\bm{\beta}\partial\bm{\beta}^{\rm T}}\bigg)\bigg|_{\bm{\beta}=\bm{\beta}^*}
\end{pmatrix}
\end{align*}
and
\begin{align*}
\tilde{\bm{\psi}}_i\frac{\partial l_i}{\partial\bm{\theta}^{\rm T}}\bigg|_{\bm{\theta}=\bm{\theta}^*}
=\begin{pmatrix}
% \displaystyle \frac{1}{2}\bigg[\frac{1}{\alpha^2}\{y_i-\bm{\beta}^{\rm T}\bm{\phi}(\bm{x}_i)\}^2-\frac{1}{\alpha}\bigg]^2
\displaystyle \frac{1}{2}\bigg(\frac{1}{\alpha^{*2}}\varepsilon_i^{*2}-\frac{1}{\alpha^*}\bigg)^2
% & \displaystyle \bigg[\frac{1}{\alpha^3}\{y_i-\bm{\beta}^{\rm T}\bm{\phi}(\bm{x}_i)\}^2-\frac{1}{\alpha^2}\bigg]\{y_i-\bm{\beta}^{\rm T}\bm{\phi}(\bm{x}_i)\}\bm{\phi}(\bm{x}_i)^{\rm T}
& \displaystyle \bigg(\frac{1}{\alpha^{*3}}\varepsilon_i^{*3}-\frac{1}{\alpha^{*2}}\varepsilon_i^*\bigg)\bm{\phi}(\bm{x}_i)^{\rm T}
\\[1.5ex]
% \displaystyle \bigg[\frac{1}{\alpha^3}\{y_i-\bm{\beta}^{\rm T}\bm{\phi}(\bm{x}_i)\}^2-\frac{1}{\alpha^2}\bigg]\{y_i-\bm{\beta}^{\rm T}\bm{\phi}(\bm{x}_i)\}\bm{\phi}(\bm{x}_i)
\displaystyle \bigg(\frac{1}{\alpha^{*3}}\varepsilon_i^{*3}-\frac{1}{\alpha^{*2}}\varepsilon_i^*\bigg)\bm{\phi}(\bm{x}_i)
% \\ \displaystyle \frac{1}{\alpha}(y_i-\bm{\beta}^{\rm T}\bm{\phi}(\bm{x}_i))\bm{\phi}(\bm{x}_i) \bigg\{\frac{2}{\alpha}(y_i-\bm{\beta}^{\rm T}\bm{\phi}(\bm{x}_i))\bm{\phi}(\bm{x}_i)^{\rm T}-2\tilde{\Lambda}\bm{\beta}-\frac{\partial\bm{\tilde{\lambda}}^{\rm T}}{\partial\bm{\beta}}({\cal B}^2\bm{1}+2\gamma_1D\tilde{\bm{\lambda}}+\gamma_2\tilde{\Lambda}^{-1}\bm{1})\bigg\}
& \displaystyle \frac{1}{\alpha^*}\varepsilon_i^*\bm{\phi}(\bm{x}_i) \bigg\{\frac{2}{\alpha^*}\varepsilon_i^*\bm{\phi}(\bm{x}_i)-\bigg(2\tilde{\bm{\Lambda}}\bm{\beta}+\frac{\partial\tilde{\bm{\lambda}}^{\rm T}}{\partial\bm{\beta}}\bm{u}\bigg)\bigg|_{\bm{\beta}=\bm{\beta}^*}\bigg\}^{\rm T}
\end{pmatrix}
\end{align*}
from (\ref{likelihood}) and (\ref{proposed_est_func_app}). 
 Similarly to the navie ridge method introduced in Section \ref{sec2}, the expectations of these quantities can be calculated to some extent. 
 Actually, we obtain the followings:
\begin{align*}
&\E\bigg(-\sum_{i=1}^n\frac{\partial\tilde{\bm{\psi}}_i}{\partial\bm{\theta}^{\rm T}}\bigg|_{\bm{\theta}=\bm{\theta}^*}\bigg)
\\
&=\begin{pmatrix}
\displaystyle \frac{n}{\alpha^{*2}}
& \displaystyle \frac{2}{\alpha^{*2}}\sum_{i=1}^n\E(\varepsilon_i^*)\bm{\phi}(\bm{x}_i)^{\rm T}
\\[1.5ex]
\displaystyle \frac{2}{\alpha^{*2}}\sum_{i=1}^n\E(\varepsilon_i^*)\bm{\phi}(\bm{x}_i)
& \displaystyle \frac{2}{\alpha^*}\sum_{i=1}^n\bm{\phi}(\bm{x}_i)\bm{\phi}(\bm{x}_i)^{\rm T}+n\bigg(2\tilde{\bm{\Lambda}}+\frac{\partial\tilde{\bm{\lambda}}^{\rm T}}{\partial\bm{\beta}}
%\bigg\{2\bigg(\frac{\partial\tilde{\bm{\lambda}}^{\rm T}}{\partial\bm{\beta}}\bigg)^{-1}{\rm diag}(\bm{\beta})+2{\rm diag}(\bm{\beta})\bigg(\frac{\partial\tilde{\bm{\lambda}}}{\partial\bm{\beta}^{\rm T}}\bigg)^{-1}+2\gamma_1D+\gamma_2\tilde{\Lambda}^{-2}\bigg\}
\bm{S}\frac{\partial\tilde{\bm{\lambda}}}{\partial\bm{\beta}^{\rm T}}
+\sum_{j=1}^p
%\bigg\{\beta_j^2+2\gamma_1(D\tilde{\bm{\lambda}})_j-\frac{\gamma_2}{\tilde{\lambda}_j}\bigg\}
t_j\frac{\partial^2\tilde{\lambda}_j}{\partial\bm{\beta}\partial\bm{\beta}^{\rm T}}\bigg)\bigg|_{\bm{\beta}=\bm{\beta}^*}
\end{pmatrix}
\end{align*}
and
\begin{align*}
\E\bigg(\sum_{i=1}^n\tilde{\bm{\psi}}_i\frac{\partial l_i}{\partial\bm{\theta}^{\rm T}}\bigg|_{\bm{\theta}=\bm{\theta}^*}\bigg)
=\begin{pmatrix}
\displaystyle \frac{1}{2\alpha^{*4}}\sum_{i=1}^n\E(\varepsilon_i^{*4})-\frac{n}{2\alpha^{*2}}
% & \displaystyle \frac{1}{\alpha^3}\E[\varepsilon_i\{y_i-\bm{\beta}^{\rm T}\bm{\phi}(\bm{x}_i)\}^2]\bm{\phi}(\bm{x}_i)^{\rm T}
& \displaystyle \sum_{i=1}^n\E\bigg(\frac{1}{\alpha^{*3}}\varepsilon_i^{*3}-\frac{1}{\alpha^{*2}}\varepsilon_i^*\bigg)\bm{\phi}(\bm{x}_i)^{\rm T}
\\[1.5ex]
% \displaystyle \frac{1}{\alpha^3}\E[\varepsilon_i\{y_i-\bm{\beta}^{\rm T}\bm{\phi}(\bm{x}_i)\}^2]\bm{\phi}(\bm{x}_i)
\displaystyle \sum_{i=1}^n\E\bigg(\frac{1}{\alpha^{*3}}\varepsilon_i^{*3}-\frac{1}{\alpha^{*2}}\varepsilon_i^*\bigg)\bm{\phi}(\bm{x}_i)
& \displaystyle \frac{2}{\alpha^{*2}}\sum_{i=1}^n\E(\varepsilon_i^2)\bm{\phi}(\bm{x}_i)\bm{\phi}(\bm{x}_i)^{\rm T}
% & \displaystyle \frac{1}{\alpha}(y_i-\bm{\beta}^{\rm T}\bm{\phi}(\bm{x}_i))\bm{\phi}(\bm{x}_i) \bigg\{\frac{2}{\alpha}(y_i-\bm{\beta}^{\rm T}\bm{\phi}(\bm{x}_i))\bm{\phi}(\bm{x}_i)^{\rm T}-2\tilde{\Lambda}\bm{\beta}-\frac{\partial\bm{\tilde{\lambda}}^{\rm T}}{\partial\bm{\beta}}({\cal B}^2\bm{1}+2\gamma_1D\tilde{\bm{\lambda}}+\gamma_2\tilde{\Lambda}^{-1}\bm{1})\bigg\}
\end{pmatrix}
\end{align*}
%$$\frac{1}{\alpha}(y_i-\bm{\beta}^{\rm T}\bm{\phi}(\bm{x}_i))\bm{\phi}(\bm{x}_i) \bigg\{\frac{2}{\alpha}(y_i-\bm{\beta}^{\rm T}\bm{\phi}(\bm{x}_i))\bm{\phi}(\bm{x}_i)^{\rm T}-2\tilde{\Lambda}\bm{\beta}-\frac{\partial\bm{\tilde{\lambda}}^{\rm T}}{\partial\bm{\beta}}({\cal B}^2\bm{1}+2\gamma_1D\tilde{\bm{\lambda}}+\gamma_2\tilde{\Lambda}^{-1}\bm{1})\bigg\}$$
from $\E\{\sum_{i=1}^n\tilde{\bm{\psi}}_i(\bm{\theta}^*)\}=\bm{0}$. 
Using these expectations and (\ref{GeneralGIC}), we can obtain an approximate GIC and provide the hyper-tuning parameters through minimizing it.

We can regard that this approximation is validated by an asymptotics if we consider the case of $(\gamma_1,\gamma_2)={\rm O}(n^{-\delta})$ for $\delta>0$. 
In this case, $\lambda_j={\rm O}(n^{-\delta})$ but $\lambda_j-\tilde{\lambda}_j={\rm O}(n^{-4\delta})$, that is, $\tilde{\lambda}_j$ can be regarded as a good approximator of $\lambda_j$.

\section{Numerical studies} \label{sec4}

\subsection{Curve fitting} \label{subsec41}

Monte Carlo simulations are conducted to investigate the effectiveness of our method. We generate random samples $\{ (x_{i},y_{i}) \mid i=1,\ldots,n\}$ from the model $y_{i} = g(x_{i}) + \varepsilon_{i}$ with a regression function $g(x)$ and errors $\{\varepsilon_{i} \mid i=1,\ldots,n\}$. We consider two simple examples: the true functions with inhomogeneous smoothness are specified by
\begin{align}
g(x)=\sin(x) + 2 \exp(-30x^2), \qquad -2 \le x \le 2,
\label{simu_1} 
\end{align}
and
\begin{align}
g(x)=\sin\{32\exp(x)^3\}, \qquad 0 \le x \le 1.
\label{simu_2}
\end{align}
Here, the function in (\ref{simu_1}), which is also used in \cite{DimGK01}, has a relatively smooth structure with a sharp peak around $x=0$, and the function in (\ref{simu_2}) has strongly and weakly smooth regions in $0\le x\le 0.5$ and $0.5\le x\le 1$, respectively.

Simulation results are obtained from 100 Monte Carlo trials, and then we evaluate mean squared errors defined by
\begin{align}
{\rm MSE} = \frac{1}{n} \sum_{i=1}^{n} \{ \hat{\bm{\beta}}^{\rm T}\bm{\phi}(x_{i}) - g(x_{i})\}^2
\label{mse}
\end{align}
to measure the goodness of fit.
It is assumed that the design points $\{x_{i} \mid i=1,\ldots,n\}$ are uniformly spaced on each domain and that the errors $\{\varepsilon_{i} \mid i=1,\ldots,n\}$ are independently and identically distributed according to ${\rm N}(0,\alpha)$.
The sample size $n$ is 50, 100, 150 or 200, and the variance of errors $\alpha$ is 0.025, 0.05, 0.075 or 0.1. 

Our method is compared with the three conventional regularization methods, the ridge (\citealt{HoeKen70}) whose tuning parameter is chosen by the GIC in (\ref{GeneralGIC}), the lasso (\citealt{Tib96}) and the adaptive lasso (\citealt{Zou06}) whose tuning parameters are chosen by the five-fold cross-validation.
The means of MSEs together with their standard deviations are reported in Tables $\ref{table_simu_1}$ and $\ref{table_simu_2}$ for the regression functions in ($\ref{simu_1}$) and ($\ref{simu_2}$), respectively.

\begin{table}[t]
\caption{Means and standard deviations of MSEs for the regression function in ($\ref{simu_1}$). }
\label{table_simu_1}
\vspace{0cm}
\begin{center}
\begin{tabular}{@{\extracolsep{\fill}}ccccccc}
\hline
 &	& 		& proposed	& ridge	& lasso & ada-lasso \\ 
\hline
& $n=$50 & mean [SD] ($\times 10^{2}$) 	& 1.08 [0.34]	& \hspace{1pt} 1.80	[0.42] & \hspace{1pt} 1.44 [0.60] & 1.26 [0.43] \\
$\alpha=$ & $n=$100	& mean [SD] ($\times 10^{3}$) 	& 6.11 [1.80]	& \hspace{1pt} 9.49 [2.06]	& \hspace{1pt} 9.11 [2.57] & 7.84 [2.43] \\
0.025 & $n=$150	& mean [SD] ($\times 10^{3}$) 	& 3.92 [1.13]	& \hspace{1pt} 6.21 [1.37]	& \hspace{1pt} 7.00 [1.87] & 5.73 [1.63] \\
& $n=$200 & mean [SD] ($\times 10^{3}$)	& 3.02 [0.86]	& \hspace{1pt} 4.51 [0.97]	& \hspace{1pt} 5.95 [1.60] & 5.02 [1.41] \\
\hline
& $n=$50 & mean [SD] ($\times 10^{2}$) 	&  1.93 [0.64]	& \hspace{1pt} 3.55	[0.84] & \hspace{1pt} 2.64 [1.11] & 2.26 [0.76] \\
$\alpha=$ & $n=$100	& mean [SD] ($\times 10^{2}$) 	& 1.09 [0.33]	& \hspace{1pt} 1.87 [0.41]	& \hspace{1pt} 1.67 [0.49] & 1.26 [0.39] \\
0.05 & $n=$150	& mean [SD] ($\times 10^{3}$) 	& 7.73 [4.96]	& 12.24 [2.71]	& 12.67 [3.65] & 9.02 [2.72] \\
& $n=$200 & mean [SD] ($\times 10^{3}$)	& 5.68 [2.98]	& \hspace{1pt} 8.87 [1.92]	& 10.62 [2.99] & 7.18 [2.09] \\
\hline
& $n=$50 & mean [SD] ($\times 10^{2}$) 	& 2.93 [1.45]	& \hspace{1pt} 5.29 [1.26]	& \hspace{1pt} 3.77 [1.59] & 3.26 [1.43] \\
$\alpha=$ & $n=$100	& mean [SD] ($\times 10^{2}$) 	& 1.73 [1.05]	& \hspace{1pt} 2.79 [0.62]	& \hspace{1pt} 2.43 [0.74] & 1.78 [0.57] \\
0.075 & $n=$150	& mean [SD] ($\times 10^{2}$) 	& 1.11 [0.63]	& \hspace{1pt} 1.82 [0.40]	& \hspace{1pt} 1.82 [0.56] & 1.21 [0.38] \\
& $n=$200 & mean [SD] ($\times 10^{3}$)	& 8.69 [5.71]	& 13.19 [2.87]	& 15.09 [4.39] & 9.58 [3.35] \\
\hline
& $n=$50 & mean [SD] ($\times 10^{2}$) 	&  4.04 [2.58]	& \hspace{1pt} 7.01	[1.67] & \hspace{1pt} 4.87 [2.00] & 4.31 [2.02] \\
$\alpha=$ & $n=$100	& mean [SD] ($\times 10^{2}$) 	& 2.17 [0.86]	& \hspace{1pt} 3.70 [0.83]	& \hspace{1pt} 3.20 [1.03] & 2.29 [0.72] \\
0.1 & $n=$150	& mean [SD] ($\times 10^{2}$) 	& 1.35 [0.38]	& \hspace{1pt} 2.41 [0.54]	& \hspace{1pt} 2.38 [0.74] & 1.52 [0.48] \\
& $n=$200 & mean [SD] ($\times 10^{2}$)	& 1.09 [0.44]	& \hspace{1pt} 1.75 [0.38]	& \hspace{1pt} 1.97 [0.59] & 1.18 [0.41] \\
\hline
\end{tabular}
\end{center}
\end{table}

\begin{table}[t]
\caption{Means and standard deviations of MSEs for the regression function in ($\ref{simu_2}$). }
\label{table_simu_2}
\begin{center}
\begin{tabular}{@{\extracolsep{\fill}}ccccccc}
\hline
 &	& 		& proposed	& ridge	& lasso & ada-lasso \\ 
\hline
& $n=$50 & mean [SD] ($\times 10^{2}$) 	&  1.32 [0.37]	& \hspace{1pt} 1.92	[0.44] & \hspace{1pt} 8.58 [12.38] & 14.79 [10.98] \\
$\alpha=$ & $n=$100	& mean [SD] ($\times 10^{3}$) 	& 7.42 [1.95]	& 10.32 [2.16]	& \hspace{1pt} 8.79 \hspace{1pt} [2.35] & 17.74 [26.95] \\
0.025 & $n=$150	& mean [SD] ($\times 10^{3}$) 	& 4.89 [1.17]	& \hspace{1pt} 6.78 [1.42]	& \hspace{1pt} 6.84 \hspace{1pt} [1.84] & \hspace{1pt} 5.95 \hspace{1pt} [3.11] \\
& $n=$200 & mean [SD] ($\times 10^{4}$)	& 2.66 [0.32]	& \hspace{1pt} 2.18 [0.25]	& \hspace{1pt} 4.90 \hspace{1pt} [0.64] & \hspace{1pt} 6.53 \hspace{1pt} [8.50] \\
\hline
& $n=$50 & mean [SD] ($\times 10^{2}$) 	&  2.64 [0.72]	& \hspace{1pt} 3.84	[0.87] & 12.07 [11.80] & 17.80 [10.14] \\
$\alpha=$ & $n=$100	& mean [SD] ($\times 10^{2}$) 	& 1.49 [0.42]	& \hspace{1pt} 2.05 [0.43]	& \hspace{1pt} 1.67 \hspace{1pt} [0.48] & \hspace{1pt} 2.54 \hspace{1pt} [2.91] \\
0.05 & $n=$150	& mean [SD] ($\times 10^{3}$) 	& 9.76 [2.44]	& 13.45 [2.84]	& 14.03 [14.28] & 10.56 \hspace{1pt} [3.91] \\
& $n=$200 & mean [SD] ($\times 10^{4}$)	& 5.25 [1.08]	& \hspace{1pt} 5.86 [1.04]	& \hspace{1pt} 8.80 \hspace{1pt} [1.65] & \hspace{1pt} 9.09 \hspace{1pt} [6.20] \\
\hline
& $n=$50 & mean [SD] ($\times 10^{2}$) 	&  4.10 [1.13]	& \hspace{1pt} 5.76	[1.31] & 16.36 [12.85] & 20.55 [10.05] \\
$\alpha=$ & $n=$100	& mean [SD] ($\times 10^{2}$) 	& 2.29 [0.61]	& \hspace{1pt} 3.07 [0.65]	& \hspace{1pt} 2.53 \hspace{1pt} [0.81] & \hspace{1pt} 3.40 \hspace{1pt} [3.31] \\
0.075 & $n=$150	& mean [SD] ($\times 10^{2}$) 	& 1.51 [0.37]	& \hspace{1pt} 2.01 [0.43]	& \hspace{1pt} 2.05 \hspace{1pt} [1.79] & \hspace{1pt} 1.53 \hspace{1pt} [0.57] \\
& $n=$200 & mean [SD] ($\times 10^{4}$)	& 9.24 [2.18]	& 11.83 [2.21]	& 15.36 \hspace{1pt} [3.26] & 13.51 \hspace{1pt} [3.33] \\
\hline
& $n=$50 & mean [SD] ($\times 10^{2}$) 	&  5.63 [1.51]	& \hspace{1pt} 7.68	[1.75] & 19.84 [11.91] & 23.46 \hspace{1pt} [9.57] \\
$\alpha=$ & $n=$100	& mean [SD] ($\times 10^{2}$) 	& 3.10 [0.82]	& \hspace{1pt} 4.09 [0.87]	& \hspace{1pt} 3.45 \hspace{1pt} [1.18] & \hspace{1pt} 4.13 \hspace{1pt} [3.60] \\
0.1 & $n=$150	& mean [SD] ($\times 10^{2}$) 	& 2.07 [0.51]	& \hspace{1pt} 2.68 [0.57]	& \hspace{1pt} 2.80 \hspace{1pt} [2.28] & \hspace{1pt} 2.05 \hspace{1pt} [1.11] \\
& $n=$200 & mean [SD] ($\times 10^{3}$)	& 1.52 [0.36]	& \hspace{1pt} 2.03 [0.39]	& \hspace{1pt} 2.51 \hspace{1pt} [0.59] & \hspace{1pt} 1.98 \hspace{1pt} [0.45] \\
\hline
\end{tabular}
\end{center}
\end{table}

\begin{figure}[t]
\begin{center}
\vspace{-10mm}
\includegraphics[width=8cm]{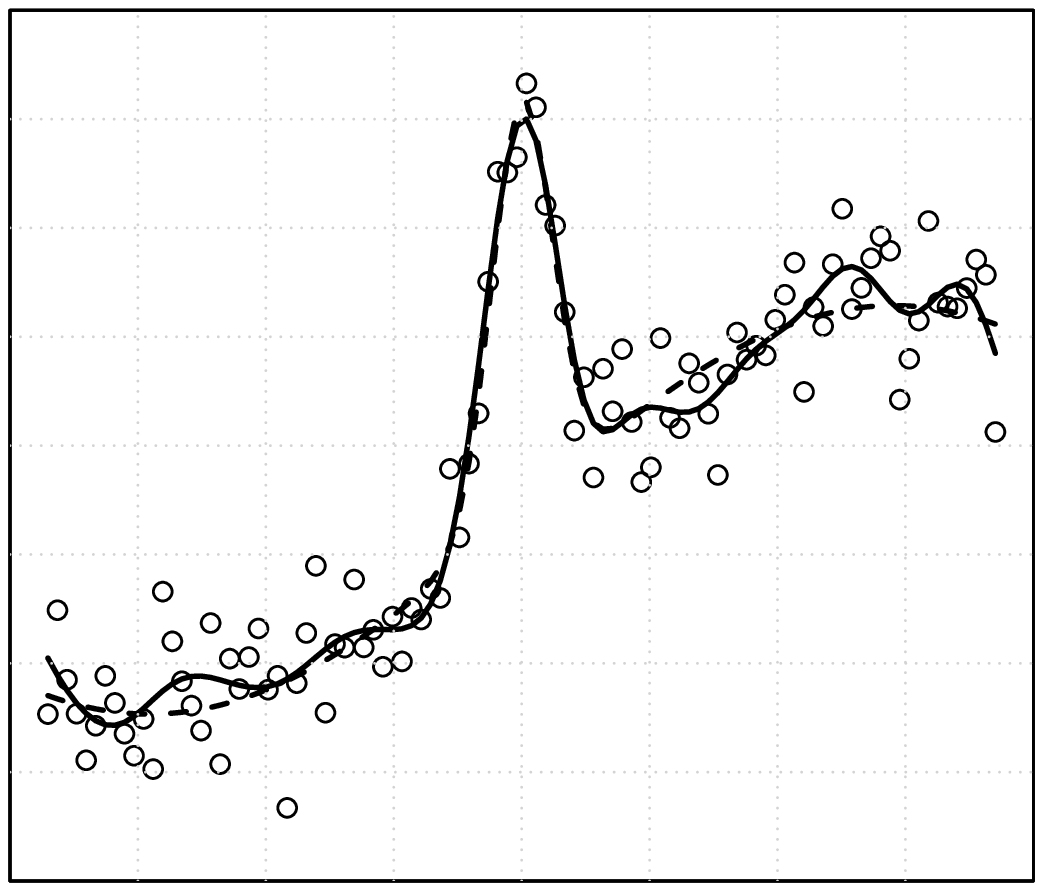}
\includegraphics[width=8cm]{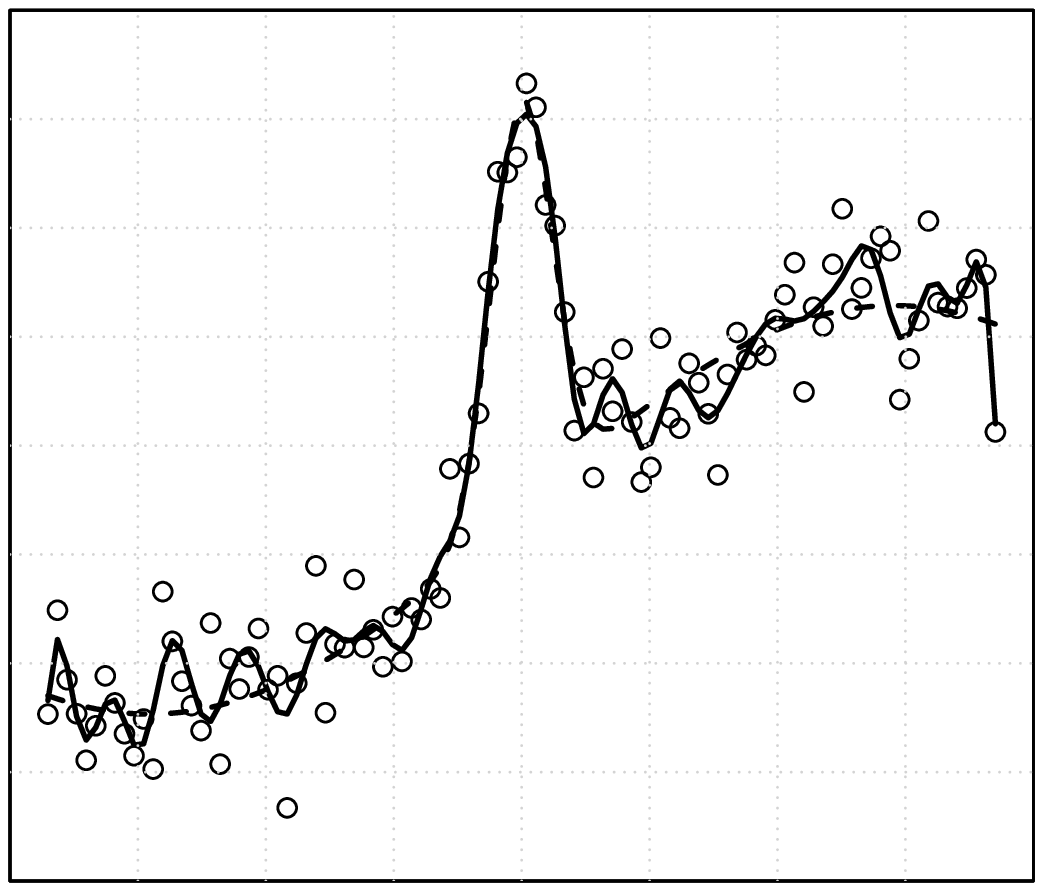}
\\[-20mm]
\includegraphics[width=8cm]{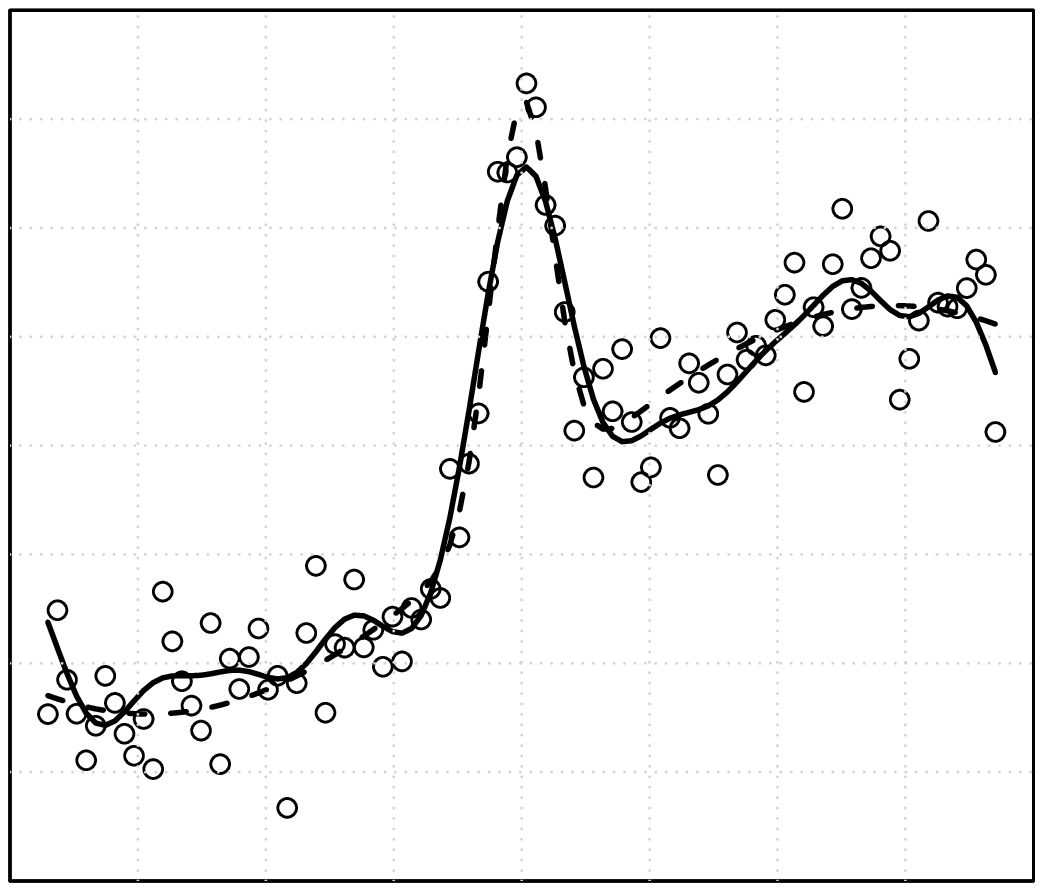}
\includegraphics[width=8cm]{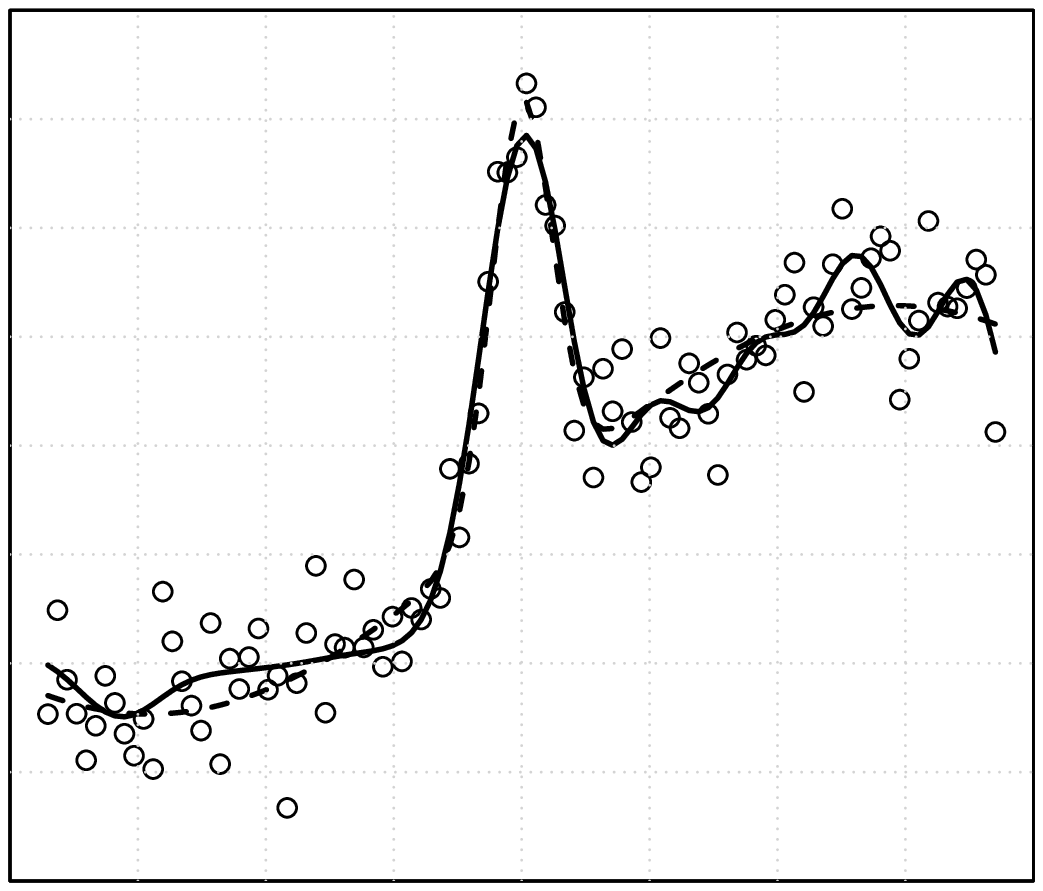}
\\[-10mm]
\caption{Estimated curves based on the proposed method (top left), ridge (top right), lasso (bottom left) and adaptive lasso (bottom right) for the regression function in ($\ref{simu_1}$).  The solid lines draw the estimated curves, and the broken lines draw the true curves.}
\label{fig_simu_1}
\end{center}
\end{figure}

\begin{figure}[t]
\begin{center}
\vspace{-10mm}
\includegraphics[width=8cm]{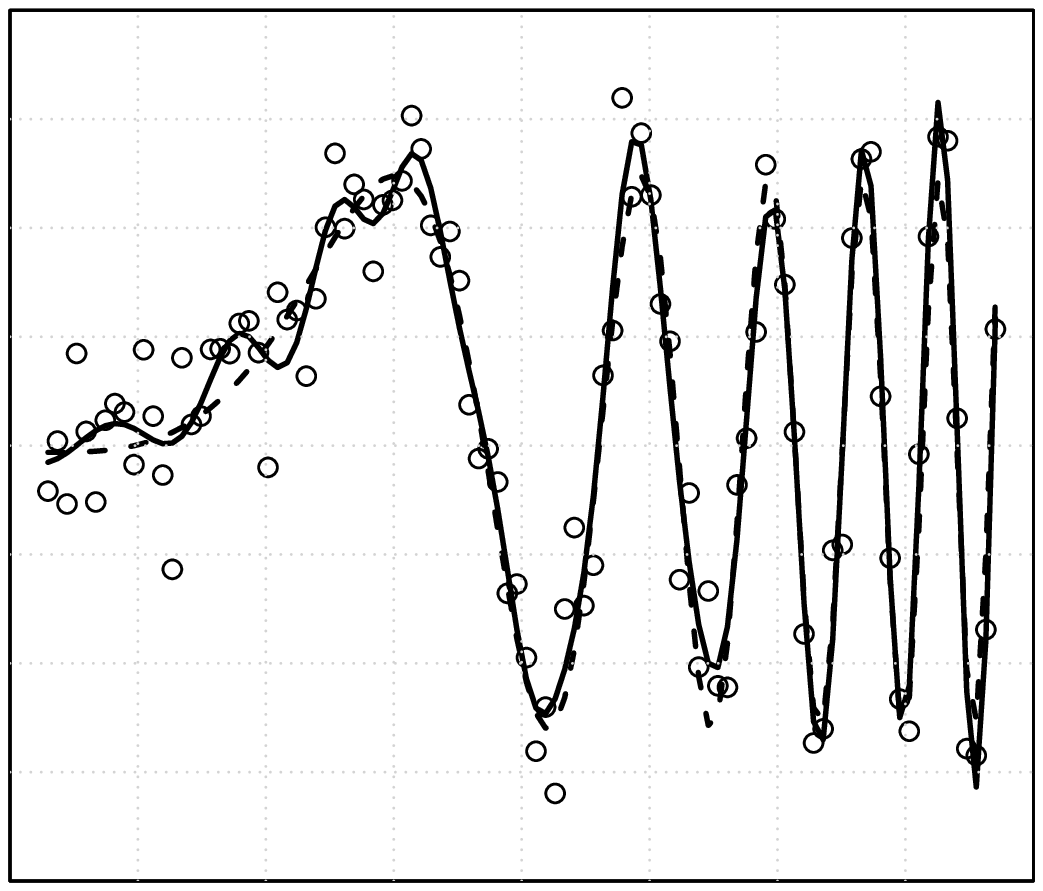}
\includegraphics[width=8cm]{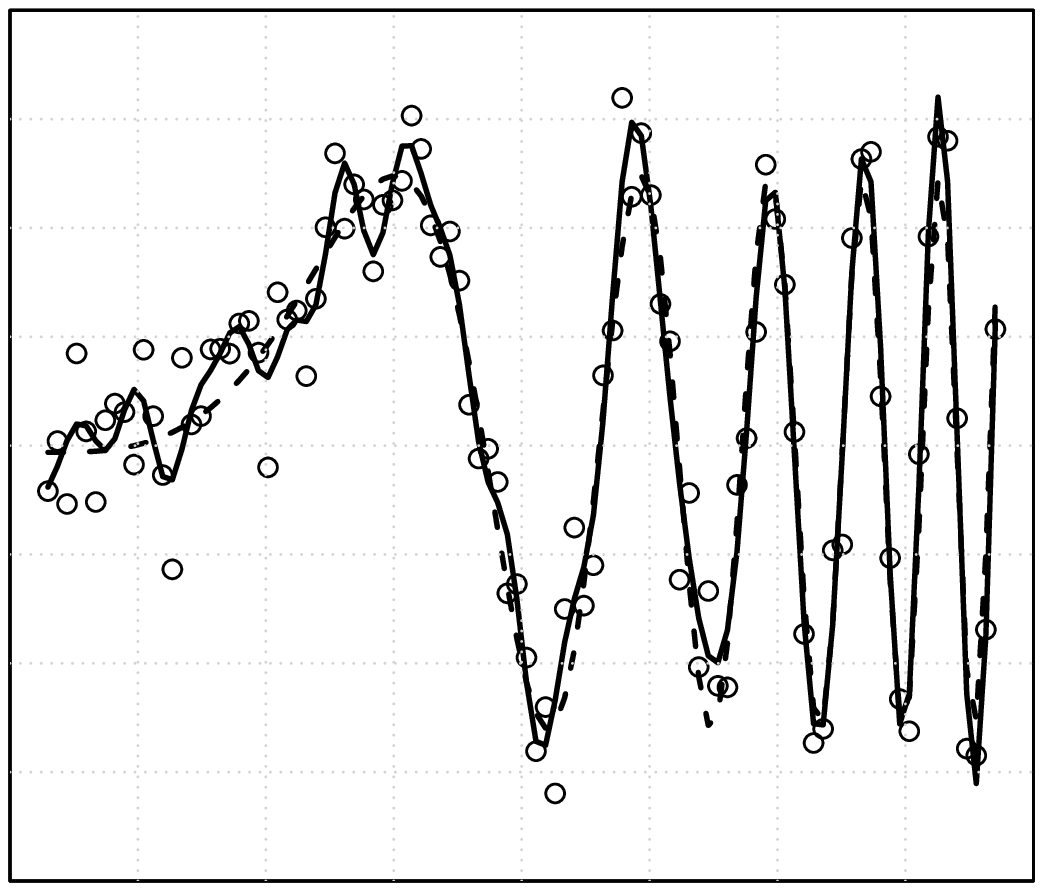}
\\[-20mm]
\includegraphics[width=8cm]{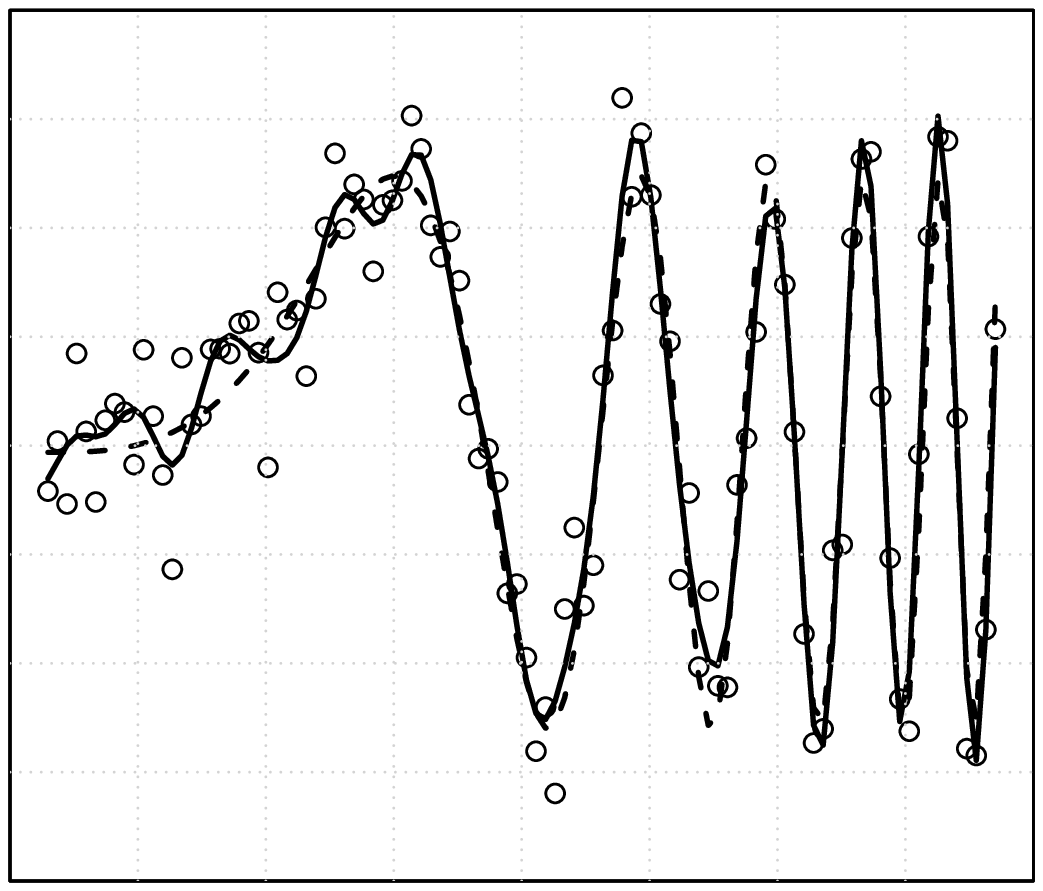}
\includegraphics[width=8cm]{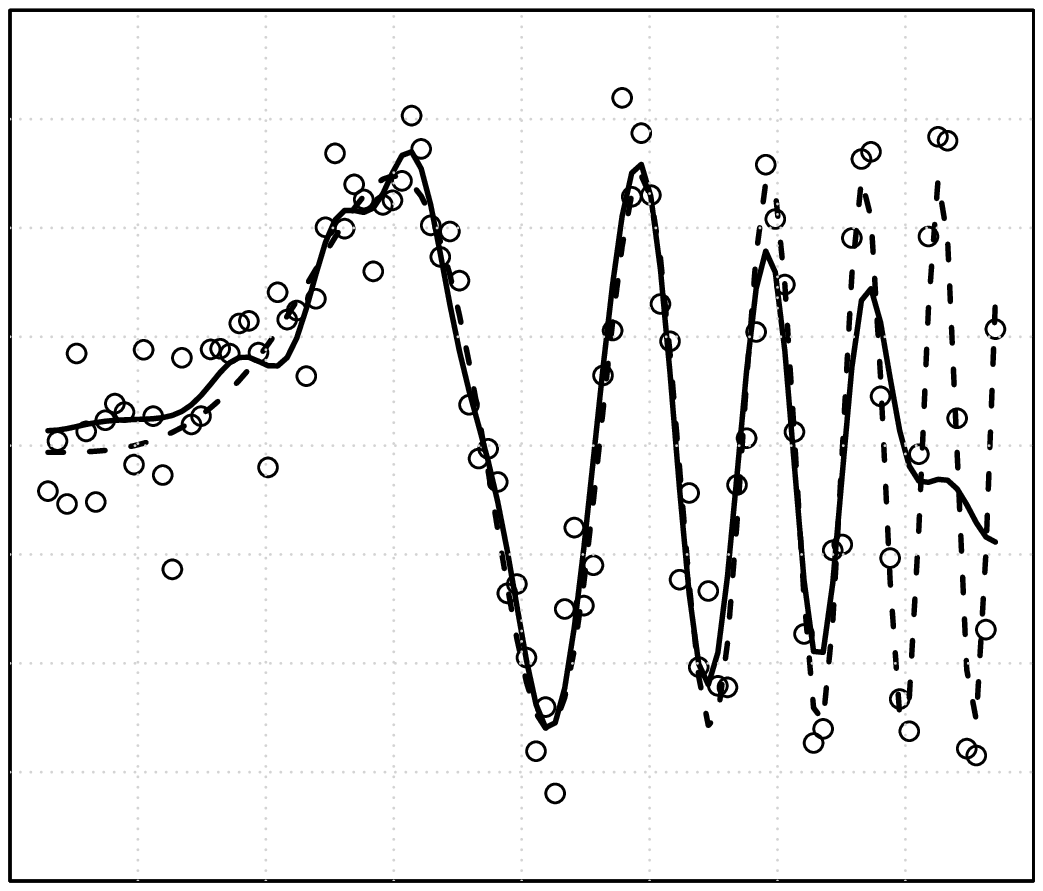}
\\[-10mm]
\caption{Estimated curves based on the proposed method (top left), ridge (top right), lasso (bottom left) and adaptive lasso (bottom right) for the regression function in ($\ref{simu_2}$).  The solid lines draw the estimated curves, and the broken lines draw the true curves.}
\label{fig_simu_2}
\end{center}
\end{figure}

For the case of $(\alpha,n)=(0.025,200)$ or $(\alpha,n)=(0.1,150)$ in Table $\ref{table_simu_1}$, the ridge or adaptive lasso is better than our method, however, for all cases except for them in Tables $\ref{table_simu_1}$ and $\ref{table_simu_2}$, our method provides smaller MSEs than the other three methods.
 Figures $\ref{fig_simu_1}$ and $\ref{fig_simu_2}$ show typical estimated curves for the case of $(\alpha,n)=(0.05,100)$ and the true regression functions in ($\ref{simu_1}$) and ($\ref{simu_2}$), respectively.
 From Figure $\ref{fig_simu_1}$, we can see that our method captures the sharp peak well and avoids over-fitting at the regions except for the peak.
 We can also see from Figure $\ref{fig_simu_2}$ that our method avoids over-fitting on $x<0.5$ and under-fitting on $x>0.5$, respectively, comparing with the other three methods.

The ridge ridge is clearly inferior to our method in Table $\ref{table_simu_1}$ and when $n$ is small in Table $\ref{table_simu_2}$.  
 The reason is that it over-fits at strongly smooth regions as seen in the top right panels of Figures $\ref{fig_simu_1}$ and $\ref{fig_simu_2}$.  
 On the other hand, it is competitive with our method when $n$ is large in Table $\ref{table_simu_2}$ because it does not cause over-fitting in this case.  
 The lasso is always clearly worse than our method in Tables $\ref{table_simu_1}$ and $\ref{table_simu_2}$.  
 Note that its MSEs are similar to those of the ridge method in Table $\ref{table_simu_1}$, but their properties are different. 
 As seen in the bottom left panel of Figure $\ref{fig_simu_1}$, the lasso causes underfitting. 
 The adaptive lasso is always superior to the ridge and lasso and comparable to our method for $\alpha=0.1$ in Table $\ref{table_simu_1}$.
 Also in Table $\ref{table_simu_2}$, it is competitive with our method for $n=150$.
 When $n$ is small in Table $\ref{table_simu_2}$, however, its MSEs are much larger than those of the other three methods.  
 It is because the adaptive lasso sometimes cannot catch the true curve at all as seen in the bottom right panel of Figure $\ref{fig_simu_2}$.

\subsection{Surface fitting} \label{subsec42}

Next we illustrate an effectiveness of our method by fitting surfaces to simulated data. 
 The random samples $\{ (\bm{x}_{i},y_{i}) \mid i=1,\ldots,n \}$ are generated from the model $y_{i} = g(x_{i1},x_{i2})+\varepsilon_{i}$. 
 We consider
\begin{align}
g(x_{i1},x_{i2})=
&\sum_{j=1}^{3}\exp[ -30 \{(x_{i1}-a_{j1})^2 + (x_{i2}-a_{j2})^2)\}]
\nonumber\\
&+\sum_{k=1}^{4}\exp[ -100 \{(x_{i1}-b_{k1})^2 + (x_{i2}-b_{k2})^2)\}]
\label{simu_surface}
\end{align}
as the true surface, where $(a_{11},a_{12})=(0.25,0.25)$, $(a_{21},a_{22})=(0.25,0.75)$, $(a_{31},a_{32})=(0.75,0.25)$, $(b_{11},b_{12})=(0.6,0.6)$, $(b_{21},b_{22})=(0.6,0.9)$, $(b_{31},b_{32})=(0.9,0.6)$ and $(b_{41},b_{42})=(0.9,0.9)$. 
 The function has strongly and weakly smooth regions in $[0,1]^2 \setminus [0.5,1]^2$ and $[0.5,1]^2$, respectively.
 Simulation results are obtained from one hundred Monte Carlo trials, and then we evaluate mean squared errors similarly to in (\ref{mse}).
 It is assumed that the design points $\{(x_{i1},x_{i2}) \mid i=1,\ldots,n\}$ are uniformly spaced on the domain $[0,1]^2$ and that the errors $\{\varepsilon_{i} \mid i=1,\ldots,n\}$ are independently and identically distributed according to ${\rm N}(0,\alpha)$.
 The sample size $n$ is 900, 1600, 2500 or 3600, and the variance of errors $\alpha$ is 0.05, 0.1, 0.15 or 0.2. 

\begin{figure}[t]
\begin{center}
\vspace{-10mm}
\includegraphics[width=9cm]{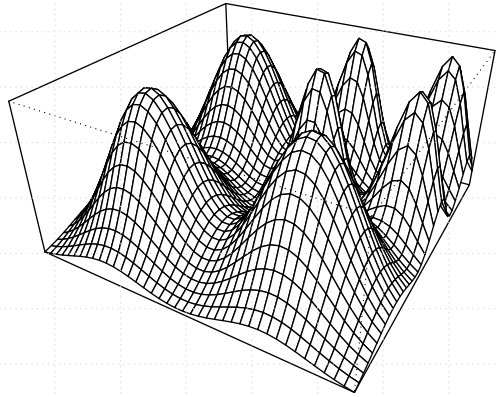}
\\[-20mm]
\caption{True surface with inhomogeneous smoothness given in ($\ref{simu_surface}$).}
\label{Fig_Truesurface}
\end{center}
\end{figure}

\begin{figure}[t!]
\begin{center}
\vspace{-10mm}
\hspace*{-10mm}
\includegraphics[width=9cm]{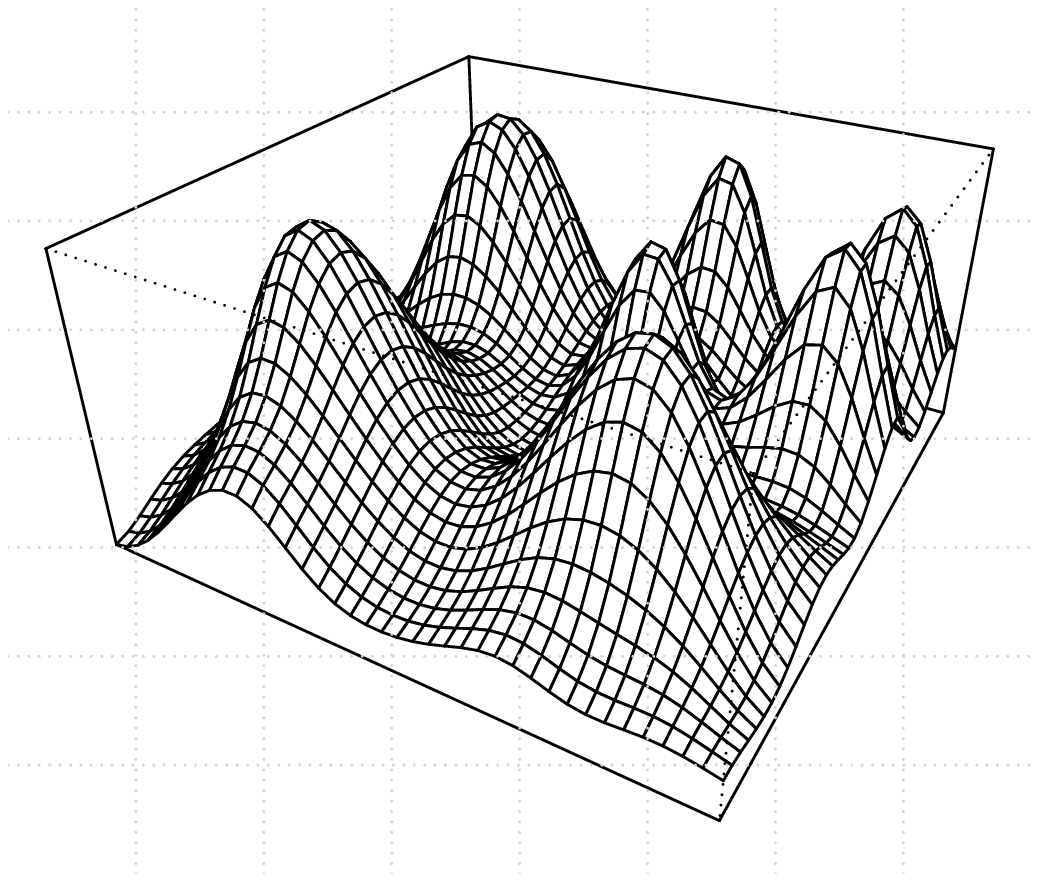}
\hspace*{-15mm}
\includegraphics[width=9cm]{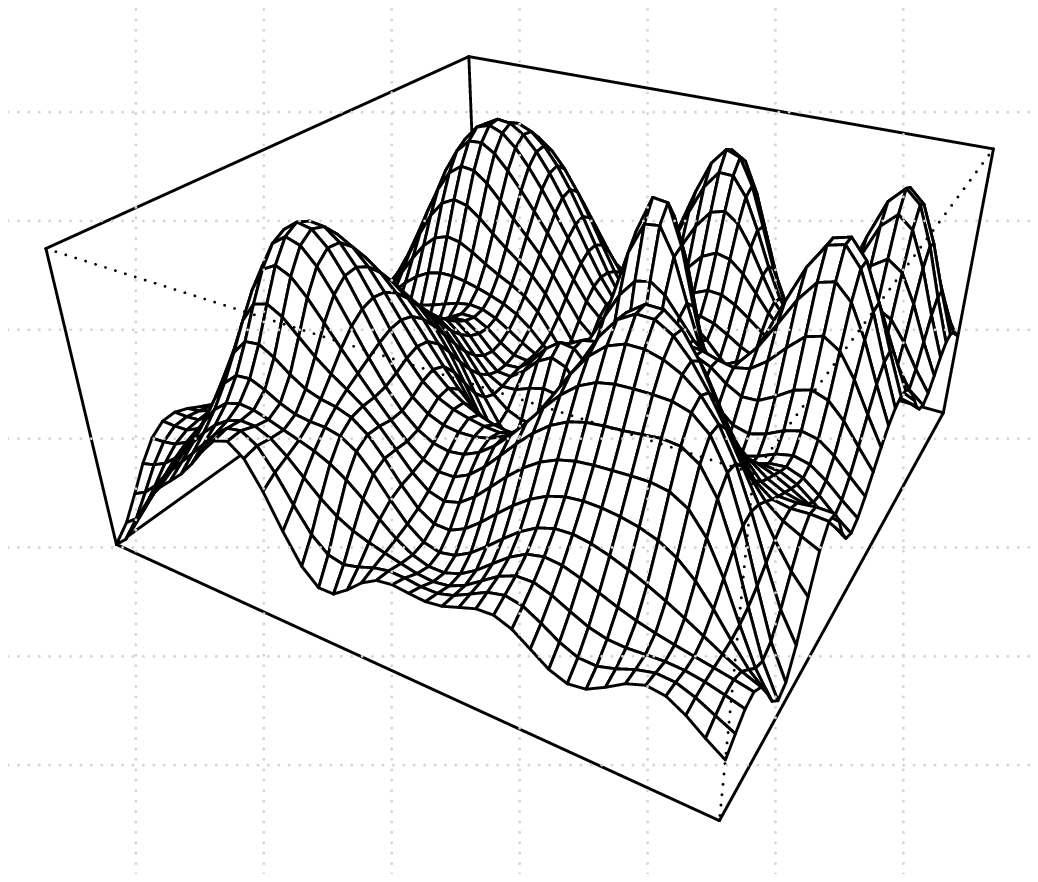}
\\[-30mm]
\hspace*{-10mm}
\includegraphics[width=9cm]{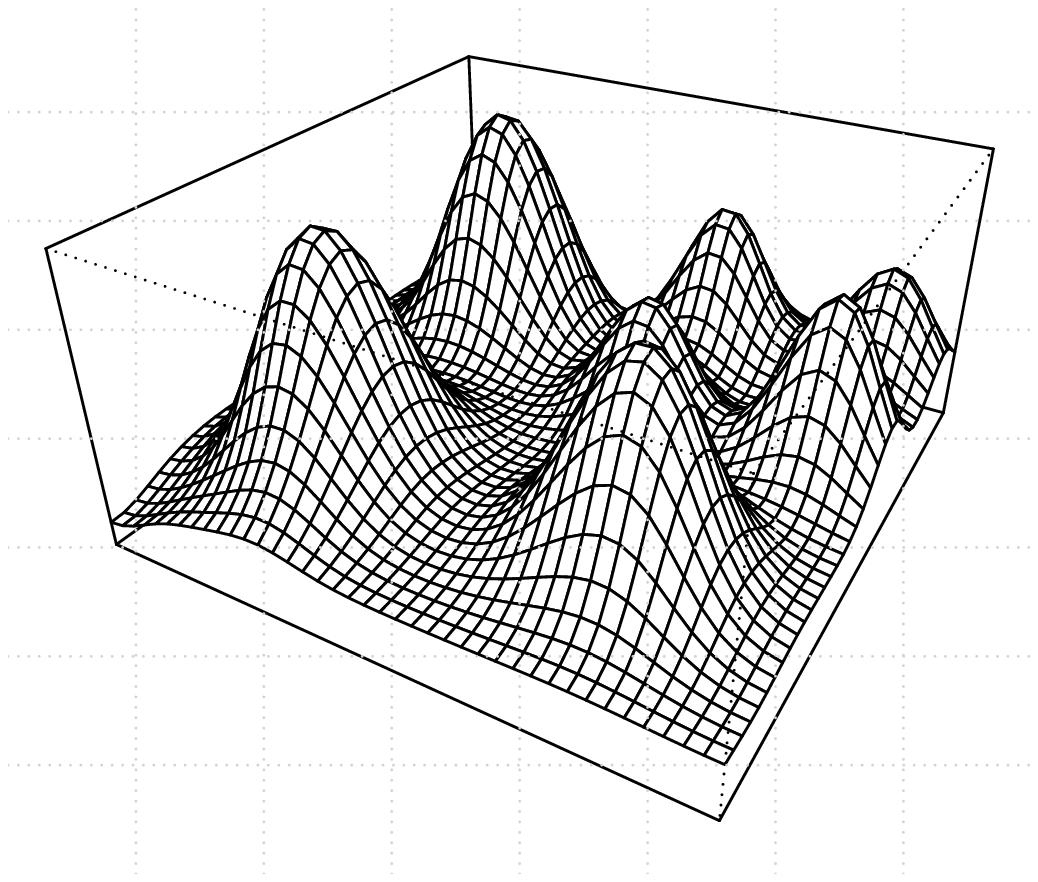}
\hspace*{-15mm}
\includegraphics[width=9cm]{Figures/s3_3.eps}
\\[-20mm]
\caption{Estimated sufraces based on the proposed method (top left), ridge (top right), lasso (bottom left) and adaptive lasso (bottom right) for the regression function in ($\ref{simu_surface}$).}
\label{Fig_surface}
\end{center}
\end{figure}

Our method is compared with the ridge choosing its tuning parameter by the GIC in (\ref{GeneralGIC}), the lasso and the adaptive lasso choosing tuning parameters by the five-fold cross-validation.
 The means of MSEs together with their standard deviations are reported in Tables $\ref{table_simu_3}$.
 Figures $\ref{Fig_Truesurface}$ shows the true surface in (\ref{simu_surface}) for the case of $(\alpha,n)=(0.1,1600)$, and its typical estimated surfaces are drawn in Figure $\ref{Fig_surface}$.

\begin{table}[t]
\caption{Mean and standard deviation of MSEs for the regression function in ($\ref{simu_surface}$). }
\label{table_simu_3}
\begin{center}
\begin{tabular}{@{\extracolsep{\fill}}ccccccc}
\hline
 &	& 		& proposed	& ridge	& lasso & ada-lasso \\ 
\hline
& $n=$900 & mean [SD] ($\times 10^{3}$) 	&  3.60 [0.68]	& \hspace{1pt} 6.67	[0.87] & \hspace{1pt} 5.84 [1.25] & \hspace{1pt} 4.60 [0.76] \\
$\alpha=$ & $n=$1600	& mean [SD] ($\times 10^{3}$) 	& 2.03 [0.31]	& \hspace{1pt} 3.76 [0.48]	& \hspace{1pt} 3.83 [0.74] & \hspace{1pt} 3.05 [0.53] \\
0.05 & $n=$2500	& mean [SD] ($\times 10^{3}$) 	& 1.35 [0.19]	& \hspace{1pt} 2.41 [0.31]	& \hspace{1pt} 2.82 [0.56] & \hspace{1pt} 2.22 [0.59] \\
& $n=$3600 & mean [SD] ($\times 10^{4}$)	& 9.78 [1.40]	& 16.54 [2.18]	& 21.69 [3.88] & 17.90 [6.16] \\
\hline
& $n=$900 & mean [SD] ($\times 10^{3}$) 	&  6.04 [1.23]	& 13.55 [1.79] & 10.42 [2.29] & \hspace{1pt} 7.60 [1.34] \\
$\alpha=$ & $n=$1600	& mean [SD] ($\times 10^{3}$) 	&  3.70 [0.56]	& \hspace{1pt} 7.51 [0.96] & \hspace{1pt} 7.21 [1.51] & \hspace{1pt} 5.02 [0.81] \\
0.1 & $n=$2500	& mean [SD] ($\times 10^{3}$) 	& 2.57 [0.38]	& \hspace{1pt} 4.82 [0.62]	& \hspace{1pt} 5.41 [1.15] & \hspace{1pt} 3.61 [0.59] \\
& $n=$3600 & mean [SD] ($\times 10^{3}$)	& 1.88 [0.28]	& \hspace{1pt} 3.31 [0.44]	& \hspace{1pt} 4.27 [0.79] & \hspace{1pt} 2.68 [0.50] \\
\hline
& $n=$900 & mean [SD] ($\times 10^{3}$) 	&  9.05 [1.78]	& 20.01	[2.61] & 15.30 [3.67] & 10.51 [1.95] \\
$\alpha=$ & $n=$1600	& mean [SD] ($\times 10^{3}$) 	& 5.53 [0.85]	& 11.27 [1.44]	& 10.32 [2.25] & \hspace{1pt} 6.78 [1.05] \\
0.15 & $n=$2500	& mean [SD] ($\times 10^{3}$) 	& 3.87 [0.59]	& \hspace{1pt} 7.24 [0.93]	& \hspace{1pt} 7.79 [1.68] & \hspace{1pt} 4.84 [0.67] \\
& $n=$3600 & mean [SD] ($\times 10^{3}$)	& 2.83 [0.42]	& \hspace{1pt} 4.96 [0.65]	& \hspace{1pt} 6.24 [1.17] & \hspace{1pt} 3.64 [0.56] \\
\hline
& $n=$900 & mean [SD] ($\times 10^{2}$) 	&  1.25 [0.27]	& \hspace{1pt} 2.67 [0.35] & \hspace{1pt} 1.99 [0.52] & \hspace{1pt} 1.32 [0.30] \\
$\alpha=$ & $n=$1600	& mean [SD] ($\times 10^{3}$) 	& 7.53 [1.22]	& 15.03 [1.93]	& 13.34 [3.01] & \hspace{1pt} 8.26 [1.26] \\
0.2 & $n=$2500	& mean [SD] ($\times 10^{3}$) 	& 5.28 [0.84]	& \hspace{1pt} 9.65 [1.24]	& 10.02 [2.23] & \hspace{1pt} 5.94 [0.89] \\
& $n=$3600 & mean [SD] ($\times 10^{3}$)	& 3.84 [0.59]	& \hspace{1pt} 6.62 [0.87]	& \hspace{1pt} 8.11 [1.54] & \hspace{1pt} 4.53 [0.72] \\
\hline
\end{tabular}
\end{center}
\end{table}

For all cases, our method provides clearly smaller MSEs than the other three methods, while the adaptive lasso is a little better than the ridge and lasso.
 From Figures $\ref{Fig_Truesurface}$ and $\ref{Fig_surface}$, we can see that the ridge causes over-fitting and that the lasso and adaptive lasso cause under-fitting. 
 On the other hand, our method captures the true surface well on the entire region, that is, avoids over-fitting on $[0,1]^2\setminus [0.5,1]^2$ and under-fitting on $[0.5,1]^2$, respectively.

\section{Real data analysis} \label{sec5}

We illustrate our procedure through analysing the earth temperature data treated in \cite{RohMJ12}.
The data $\{ (x_{i},y_{i}) \mid i=1,\ldots,3109 \}$ consist of monthly the averages of earth temperatures in the period from 1753 to 2011, where $x_i$ is the $i$-th time value and $y_i$ is the temperature at $x_i$.

\begin{figure}[t]
\begin{center}
\includegraphics[width=8cm]{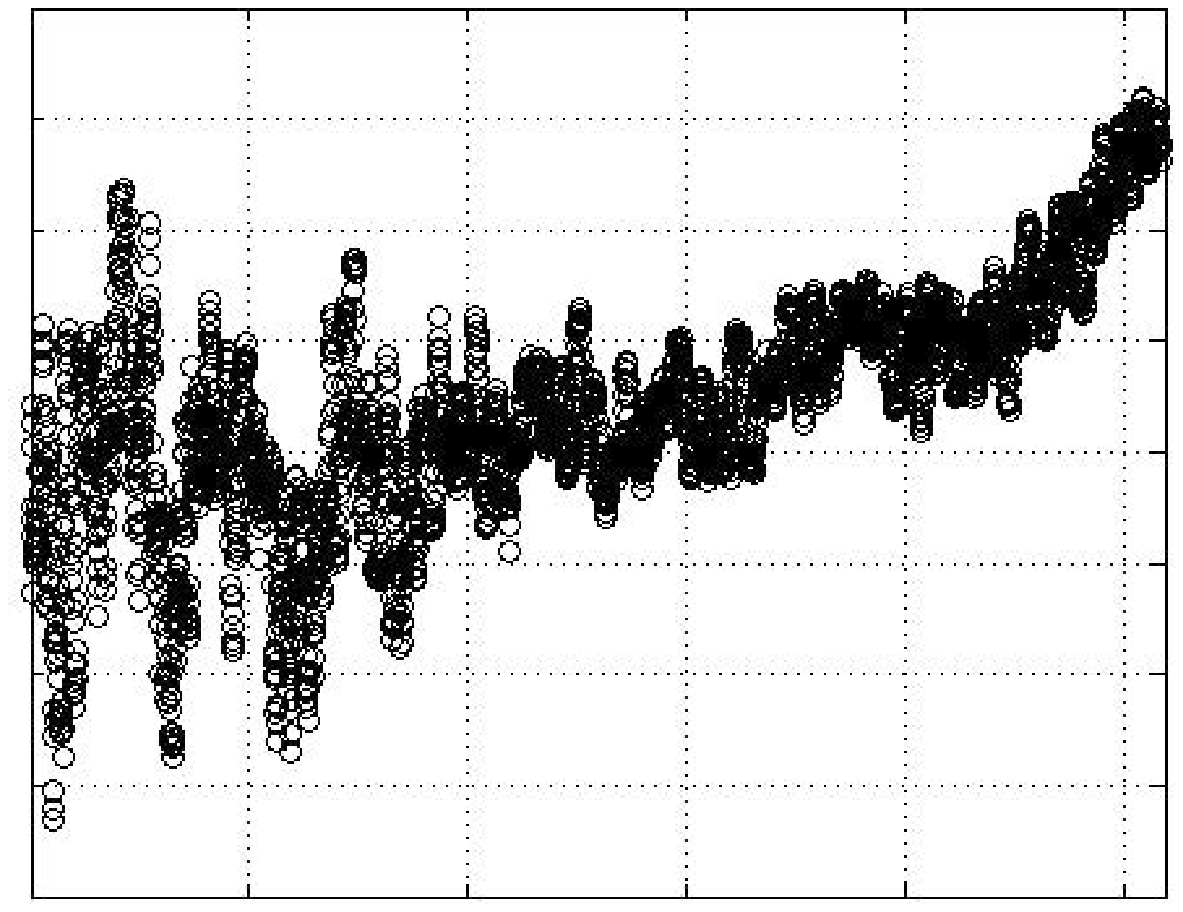}
\\[-2mm]
\includegraphics[width=8cm]{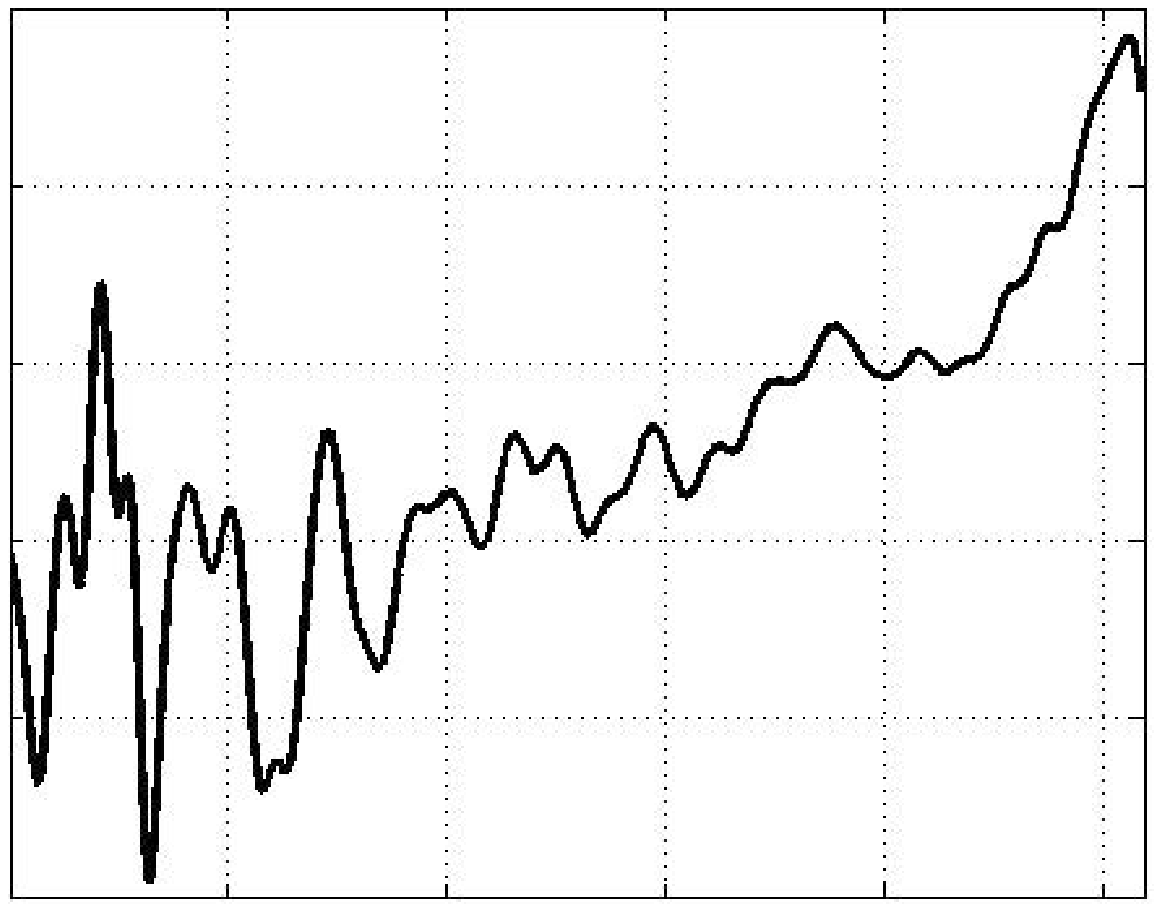}
\includegraphics[width=8cm]{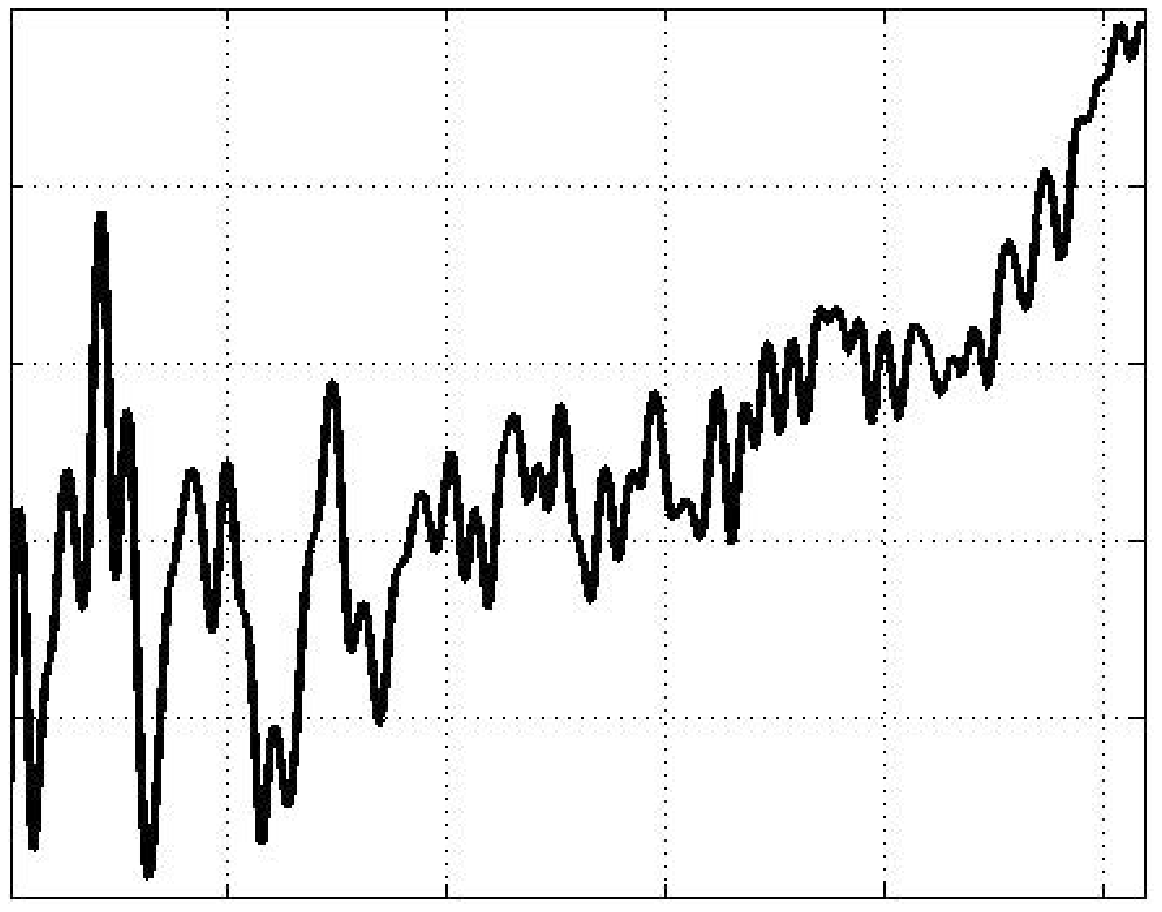}
\\[-6mm]
\label{$earth_data$}
\caption{The earth temperature data (top) and estimated curves with proposed method (bottom left) and ridge (bottom right).}
\label{Fig_earth}
\end{center}
\end{figure}

First, we applied the ridge method with choosing the tuning parameter by the GIC in (\ref{GeneralGIC}).
 The data and the estemated curve are reported in Figure $\ref{Fig_earth}$. 
 Of course, we do not know its true curve, but it seems that the ridge method yields over-fitting in the period from 1850 to 2000, as seen in our simulation studies.
 To avoid the over-fitting, next, we applied our method with choosing the hyper-tuning parameters by the approximated GIC derived in Section \ref{subsec33}.
 The result is also reported in Figure $\ref{Fig_earth}$.
 Our method clearly removes the fluctuation in the period in which over-fitting is observed, and therefore we can say that the difference between our and the ridge methods is not ignorable.

\section{Concluding remarks} \label{sec6}

We have proposed a new efficient nonlinear regression modeling with smoothly varying regularization method. The main substantive contributions of this work is to introduce the varying degree of smoothness method in constructing the nonlinear regression modeling via regularization with adaptive-type penalties. Determining a value of a tuning parameter and estimating unknown parameters are implemented by maximizing the proposed regularized log-likelihood function. The estimated curves based on our method tend to be flexible on the weakly smooth region and smooth on the strongly smooth region. To obtain appropriate values of tuning parameters, we also propose a model selection criterion for evaluating constructed models from an information-theoretic point of view. Our method is applied to the analysis of simulated data and Earth surface temperature data. The conventional method causes over-fitting and under-fitting on strongly and weakly smooth regions, respectively, but our method considerably reduces this problem. From these results, we conclude that our method is effective in the case that the underlying function has inhomogeneous smoothness.

In this paper, we focus on the ridge method as the regularization. 
 Recently, as can be seen from the publication of the special issue edited by \cite{Jos20} in Technometrics, the ridge method recaptures the spotlight.  
 For example, in the over-parametrized linear regression framework with ``double descent'' behavior of the prediction risk, which is observed also in deep neural networks, \cite{HasMRT19} showed that the minimum $\ell_2$ norm least squares estimator achieved a good generalization despite having zero training error, but the optimally-tuned ridge estimator additionally dominated it in risk.  
 They derive an asymptotic property of the ridge estimator in a high-dimensional setting similarly to in \cite{Dic16} and \cite{DobWag18}. 
 Close relationship between the ridge regularization and early stopped gradient descent is another interesting recent topic (\citealt{WeiYW17}, \citealt{AliKT19}). 
 Our method can be regarded as a development of the ridge one.
 To investigate an asymptotic property in a high-dimensional setting including the over-parametrized case and to construct a relationship with the early stopped gradient descent is our future theme.

\section*{Acknowledgment}

SK was supported by JSPS KAKENHI Grant Numbers JP19K11854 and JP20H02227, and YN was supported by JSPS KAKENHI Grant Numbers JP16K00050.

\bibliography{myrefs}

\end{document}